\documentclass[epj]{svjour}
\usepackage{graphics}
\usepackage{cite}

\begin{document}
\title{Investigating the topology of interacting networks}
\subtitle{Theory and application to coupled climate subnetworks}
\author{J.F. Donges\inst{1,2} \and H.C.H. Schultz\inst{1,3} \and N. Marwan\inst{1} \and Y. Zou\inst{1} \and J. Kurths\inst{1,2}
}                     
%
%
\institute{Potsdam Institute for Climate Impact Research, P.O.Box~60\,12\,03, 14412 Potsdam, Germany\\ (\email{donges@pik-potsdam.de}) \and Department of Physics, Humboldt University of Berlin, Newtonstr.~15, 12489 Berlin, Germany \and Department of Physics, Free University Berlin, Arnimallee~14, 14195 Berlin, Germany}
%
\date{Draft: \today / Received: date / Revised version: date}
%
\abstract{
Network theory provides various tools for investigating the structural or functional topology of many complex systems found in nature, technology and society. Nevertheless, it has recently been realised that a considerable number of systems of interest should be treated, more appropriately, as interacting networks or networks of networks. Here we introduce a novel graph-theoretical framework for studying the interaction structure between subnetworks embedded within a complex network of networks. This framework allows us to quantify the structural role of single vertices or whole subnetworks with respect to the interaction of a pair of subnetworks on local, mesoscopic and global topological scales. \\Climate networks have recently been shown to be a powerful tool for the analysis of climatological data. Applying the general framework for studying interacting networks, we introduce coupled climate subnetworks to represent and investigate the topology of statistical relationships between the fields of distinct climatological variables. Using coupled climate subnetworks to investigate the terrestrial atmosphere's three-dimensional geopotential height field uncovers known as well as interesting novel features of the atmosphere's vertical stratification and general circulation. Specifically, the new measure ``cross-betweenness" identifies regions which are particularly important for mediating vertical wind field interactions. The promising results obtained by following the coupled climate subnetwork approach present a first step towards an improved understanding of the Earth system and its complex interacting components from a network perspective.
} 
\maketitle
%
%
\section{Introduction}
\label{sec:intro}

Complex networks are recognised as a structurally simple, yet powerful representation of the manifold systems of intricately interacting elements found in nature, technology and human society \cite{Albert2002,Newman2003,Boccaletti2006}. Drawing on ideas from mathematical graph theory and statistical physics, complex network theory allows a detailed and quantitative investigation of the interaction topology of networked systems \cite{Costa2005} as well as exploring the interplay between network structure and dynamics on the interacting elements \cite{Arenas2008}. 

\begin{figure}[th!]
\centering
\resizebox{1.0\columnwidth}{!}{%
  \includegraphics{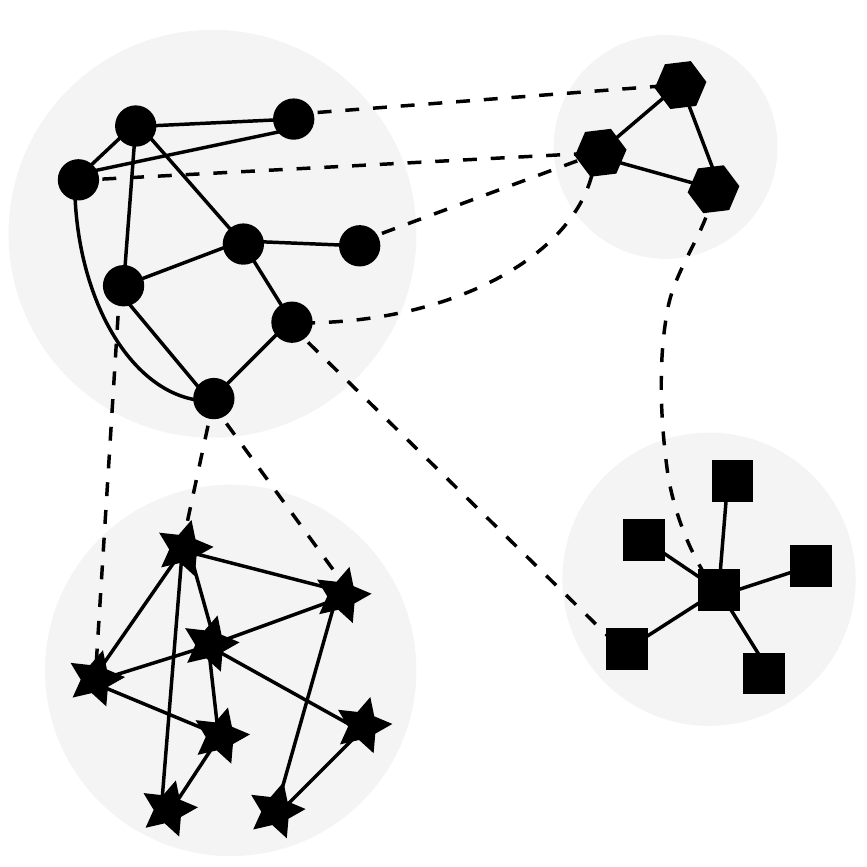}
}
\caption{Systems of interacting networks or networks of networks are a natural representation of many systems found in nature, technology and society. The partition into interacting subnetworks (indicated by vertex symbols of different shape) is either naturally induced by the considered problem, e.g., consider interdependent infrastructures \cite{Vespignani2010,Parshani2010,Buldyrev2010} or cortical areas of the mammalian brain \cite{Zhou2006,Zhou2007}, or may be generated by a community detection algorithm \cite{Fortunato2010}. Such a system is characterised by dependencies within subnetworks (internal edges, continuous lines) as well as interactions between different subnetworks (cross edges, dashes lines).}
\label{fig:interacting_networks_sketch}
\end{figure}

Most studies have so far concentrated on networks where vertices represent single elements or subsystems, and edges indicate interactions or relationships between vertices. However, it has recently been realised that a considerably large class of systems of interest warrants a more natural representation as interacting networks or networks of networks for an appropriate description of their interaction structure (Fig. \ref{fig:interacting_networks_sketch}). Notable examples are representations of the mammalian cortex, where cortical areas form complex subnetworks that are themselves linked via a complex network topology \cite{Zhou2006,Zhou2007}, systems of interacting populations of heterogeneous oscillators \cite{Barreto2008,So2008}, or mutually interdependent infrastructure networks \cite{Vespignani2010,Parshani2010} such as the power grid and electricity consuming communication networks \cite{Buldyrev2010}. More generally, and particularly if the system's representation as an interacting network is a less obvious choice than for the aforementioned examples, networks with a pronounced community structure may be viewed as interacting networks, where subnetworks are constituted by communities or clusters as identified by some community detection algorithm \cite{Fortunato2010}. A related but distinct concept is presented by layered networks \cite{Kurant2006a,Kurant2006b,Kurant2007}, where essentially different sets of edges (layers) are considered connecting the same substrate of vertices, e.g., describing a two-layered transportation network with roads forming a layer of physical connections between locations and a superposed layer of virtual connections induced by actual pathways of traffic flow on the physical layer.

Recently, climate networks representing the statistical similarity structure of a spatio-temporally resolved climatological field were successfully employed for revealing novel aspects of climate dynamics in reanalysis data sets and in the results of global climate models \cite{Donges2009,Donges2009a,Donner2008,Gozolchiani2008,Tsonis2004,Tsonis2008a,Tsonis2008b,Yamasaki2008}. These findings include among others insights into the effect of the El-Ni\~no Southern Oscillation (ENSO) on the global correlation structure of several climatological observables \cite{Tsonis2008a,Yamasaki2008} as well as the detection of a backbone of significantly increased matter and energy flow in the global surface air temperature field \cite{Donges2009,Donges2009a}. Now understanding the complex interactions between different domains of the Earth system, which may themselves be viewed as complex dynamical systems, e.g., the atmosphere, hydrosphere, cryosphere and biosphere, remains a great challenge for modern science \cite{Schellnhuber1999}. The urge to make progress in this field is particularly pressing as substantial and mutually interacting components of the Earth system (tipping elements), such as the Indian Monsoon and ENSO, may soon pass a bifurcation point (tipping point) due to global climate change and consequently experience abrupt and possibly irreversible transitions in their dynamics and function \cite{Lenton2008}. Mapping the complex interdependency structure of subsystems, components or processes of the Earth system to a network of interacting networks provides a natural, simplified and condensed mathematical representation. This structure could in turn be harnessed for generating new insights by graph-theoretical analysis and, hence, fostering an improved understanding of the Earth system's vulnerability to perturbations like anthropogenic emissions of greenhouse gases \cite{Donner2009}. As a first step in this direction, in this work we develop a novel approach termed \emph{coupled climate subnetwork analysis} for representing and studying the statistical relationships between several fields of climatological observables and apply it to investigate the atmosphere's vertical dynamical structure and general circulation \cite{Schneider2006,Hartmann2007} from a network perspective.

So far research on interacting networks has focussed on global properties of these systems, e.g., percolation thresholds \cite{Buldyrev2010}. However, when dealing with networks of networks it is of major interest to determine in detail the importance and role of single vertices for the interaction or communication between different subnetworks as well as to characterise their mutual interaction topology, for example for studying the vulnerability of coupled and interdependent networked systems to random perturbations or targeted attacks. Here we propose a general and novel framework that allows to quantitatively investigate the interaction structure of networks of networks on different topological scales. We derive graph-theoretical measures that allow to answer questions like: Does a vertex have a large direct influence on and/or is it an efficient transmitter of information to a specific subnetwork? Which amount of control does a vertex have on the interaction between two subnetworks? Are two subnetworks topologically well separated or tightly intertwined and is their interaction structure well organised or random? The methodology introduced in this work therefore creates a new setting for a detailed graph-theoretical assessment of the functional roles of vertices within complex networks of networks, e.g., integration or segregation of information in corticocortical networks of cats or macaque monkeys \cite{Zamora2009,Zamora2010}.

After introducing measures and related theoretical considerations for characterising the topology of interacting networks in Sect.~\ref{sec:theory}, we employ this framework in Sect.~\ref{sec:application} for studying the atmosphere's vertical dynamical structure within the context of climate network analysis. Conclusions are drawn in Sect.~\ref{sec:conclusion}.

%
%
\section{Theory: The topology of interacting networks}
\label{sec:theory}

Consider a network $G=(V,E)$, where $V=\left\{1,..,N\right\}$ is a set of vertices or elements and $E$ a set of edges or interactions with $N=|V|$. As we wish to study a network of interacting subnetworks, we consider a decomposition of the vertex set $V$ into disjoint sets $V_i$ such that $\cup_i V_i = V$ and $V_i \cap V_j = \emptyset, \forall i \neq j$, where the number of vertices in subset $V_i$ is $N_i=|V_i|$. Similarly, the edge set $E$ is decomposed into sets $E_{ii}$ containing edges between vertices inside $V_i$ and sets $E_{ij}$ of edges connecting vertices from $V_i$ and $V_j$, i.e., $\cup_{ij} E_{ij} = E$ with $E_{ij} \cap E_{kl} = \emptyset, \forall (i,j) \neq (k,l)$. In other words, the mutual interactions between subnetworks $G_i=(V_i,E_{ii})$ ($G_i$ is the induced subgraph of $V_i$) are described by the edge sets $E_{ij}$ for $i \neq j$. For simplicity we restrict ourselves to undirected and unweighted simple graphs, since the generalisation of the concepts and measures introduced below is straightforward. This type of network is conveniently represented by the symmetric adjacency matrix $A_{pq}$ with $A_{pq}=1$ if $\{p,q\} \in E$ and $A_{pq}=0$ otherwise. In this work, indices $i,j,k,l$ always denote subnetworks while $v,w,p,q$ designate single vertices. 

In the following we define several local as well as global network measures to quantify and investigate the interaction topology of networks of networks on different topological scales. Here, local network measures assign a real number to a vertex $v \in V_i$ in relation to (a generally different) subnetwork $G_j$, or to any other vertex $w \in V$ depending on two subnetworks $G_i, G_j$. They are inspired by the ``trinity" of classical and frequently used centrality measures degree, closeness and betweenness \cite{Freeman1979,Newman2003,Boccaletti2006}, as well as the local clustering coefficient \cite{Watts1998}, and accordingly quantify direct influence on $G_j$ (cross-degree, Eq. (\ref{eq:cross_degree})), local organisation of interdependency with $G_j$ (local cross-clustering, Eq. (\ref{eq:local_cross_clustering})), efficiency of interaction with $G_j$ (cross-closeness, Eq. (\ref{eq:cross_closeness})) and the control over communication between $G_i$ and $G_j$ (cross-betweenness, Eq. (\ref{eq:cross_betweenness})), respectively. The global network measures we introduce assign a real number to a pair of subnetworks $(G_i,G_j)$. They are derived from the well established measures edge density, global clustering coefficient and average path length \cite{Watts1998}. These global network measures quantify various overall aspects of the interaction between two subnetworks such as its degree of organisation (global cross-clustering coefficient and cross-transitivity, Eqs. (\ref{eq:cross_clustering},\ref{eq:cross_transitivity})), efficiency and speed of information transfer (cross-average path length, Eq. (\ref{eq:cross_average_path_length})) or their mutual interconnectivity (cross-edge density, Eq. (\ref{eq:cross_edge_density})).

Along these lines, similar generalisations of other local and global network properties (like local random walk betweenness or global efficiency among many others \cite{Albert2002,Newman2003,Boccaletti2006}) may be derived to quantify additional nuances of the topology of interacting networks. Adaptations for directed and edge- or vertex-weighted networks \cite{Heitzig2010} are also straightforward to deduce. Similarly it is possible on the basis of our proposed framework to design measures to take into account different qualities or functions of vertices and edges within or between subnetworks, e.g., the additional constraint that the functioning of vertices $v \in V_i$ depends on the functioning of vertices $v' \in V_j$ studied by Buldyrev \textit{et al.} \cite{Buldyrev2010}. The selection of measures presented here was chosen to be as concise as possible, while at the same time representing all classes of commonly used network quantifiers. 

In the following definitions, we always assume $v \in V_i$ if not indicated otherwise. The formula explicitely account for the general case $i \neq j$ but can be, nevertheless, easily modified to suit the special case $i = j$. Furthermore, the term \emph{cross} generally relates to the interaction between subnetworks $G_i,G_j$, whereas \emph{internal} refers to the structure within a single subnetwork.

\subsection{Local measures}

\paragraph{Cross-degree centrality}

$k_v^{ij}$ gives the number of neighbours of the vertex $v$ within subnetwork $G_j$,
\begin{equation}
k_v^{ij} = k_v^j = \sum_{q \in V_j} A_{vq}, v \in V_i. \label{eq:cross_degree}
\end{equation}
This measure thus captures the importance of $v$ for the interaction or communication between two subnetworks $G_i, G_j$ in terms of the number of direct connections it projects between $G_i$ and $G_j$. For brevity, we will in the following suppress the somewhat redundant index $i$ whenever possible, e.g., write $k_v^j$ instead of $k_v^{ij}$.

The standard degree centrality $k_v$ considering the full network $G$ can be obtained by summing up the contributions from all subnetworks:
\begin{equation}
k_v = \sum_j k_v^j. \label{eq:degree}
\end{equation}

\paragraph{Local cross-clustering coefficient}

\begin{figure}[t]
\centering
\resizebox{1.0\columnwidth}{!}{%
  \includegraphics{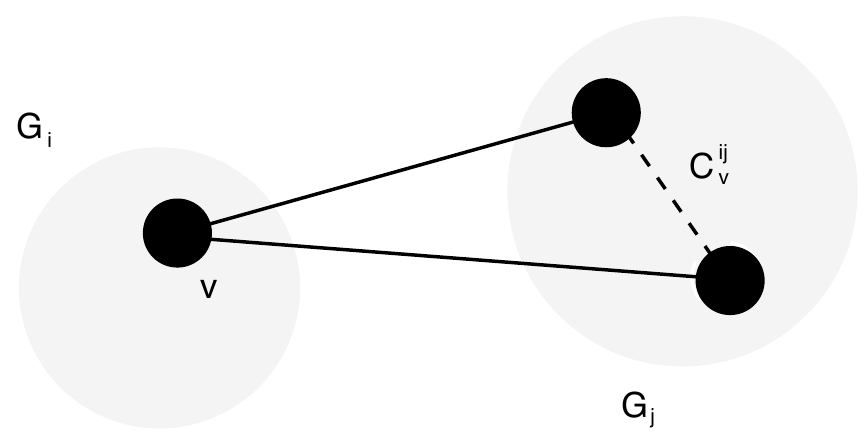}
}
\caption{The local cross-clustering coefficient $\mathcal{C}_v^{ij}$ is the probability that two randomly drawn neighbours of vertex $v$ from subnetwork $G_j$ are neighbours themselves, where $v$ belongs to subnetwork $G_i$.}
\label{fig:cross_clustering_sketch}
\end{figure}

$\mathcal{C}_v^{ij}$ estimates the probability that two randomly drawn neighbours of $v$ from subnetwork $G_j$ are also neighbours (Fig. \ref{fig:cross_clustering_sketch}),
\begin{equation}
\mathcal{C}_v^{ij} = \mathcal{C}_v^{j} = \frac{1}{k_v^{j}(k_v^{j} - 1)} \sum_{p \neq q \in V_j} A_{vp} A_{pq} A_{qv}. \label{eq:local_cross_clustering}
\end{equation}
For all vertices $v^\ast$ with $k_{v^\ast}^j \in \left\{0,1\right\}$ we set $\mathcal{C}_{v^\ast}^{j}=0$. $\mathcal{C}_v^{j}$ quantifies the tendency of vertices to form clusters spanning two subnetworks and therefore contains important information on the interaction structure between them. Assuming no correlations between the occurrence of edges within $G_j$ and between $G_i$ and $G_j$ (described by the sets $E_{jj}$ and $E_{ij}$, respectively) the expectation value for $\mathcal{C}_v^{j}$ is the internal edge density of subnetwork $G_j$, i.e., $\left<\mathcal{C}_v^{j}\right> = \rho_j, \forall v$, where $\rho_j= 2 |E_{jj}| / (N_j (N_j - 1))$. This lack of correlations would arise if edges in $E_{ij}$ were distributed randomly and independently between subnetworks $V_i$ and $V_j$. Hence, $\mathcal{C}_v^{j} \gg \rho_j$  or $\mathcal{C}_v^{j} \ll \rho_j$ indicates significant (anti-) correlations in the connectivity between both subnetworks pointing at design principles or details of growth processes depending on the specific application (e.g., see Sect.~\ref{sec:results}).

Compared to the other local measures introduced in this section, $\mathcal{C}_v^{j}$ has a less direct relationship with the standard local clustering coefficient $\mathcal{C}_v$ \cite{Newman2003}, since the contributions from triangles containing vertices from three different subnetworks also have to be taken into account (second summand in Eq. (\ref{eq:clustering_decomp})),
\begin{eqnarray} \label{eq:clustering_decomp}
\mathcal{C}_v &=& \frac{1}{k_v (k_v - 1)} \left( \sum_j k_v^j (k_v^j - 1) \mathcal{C}_v^{j} \right. \\ \nonumber
&& \left. + \sum_{k \neq l \neq i} \sum_{p \in V_k, q \in V_l} A_{vp} A_{pq} A_{qv} \right),
\end{eqnarray}
where again this holds only if $k_v>1$.

\paragraph{Cross-closeness centrality}

$c_v^{ij}$ measures the topological closeness of $v$ to subnetwork $G_j$ along shortest paths,
\begin{equation}
c_v^{ij} = c_v^j = \frac{N_j}{\sum_{q \in V_j} d_{vq}}, \label{eq:cross_closeness}
\end{equation}
where $d_{vq}$ is the shortest path length between vertices $v$ and $q$. If no path exists, i.e., the vertices are not mutually reachable, $d_{vq}=N-1$ is set as an upper bound, since this is the longest possible path length in the considered network. It is important to note that for generality, we do not restrict paths to subnetworks $G_i, G_j$ or a particular order of vertices as in earlier works \cite{Flom2004}, which might however be appropriate depending on the specific application. In contrast, the shortest paths analysed here and below may contain any vertices $w \in V$ in any order depending on the topology of the full graph $G$.

Cross-closeness therefore quantifies the efficiency of interaction between a particular vertex and a specific subnetwork. A vertex with high cross-closeness is likely to be important for the fast exchange of information with a certain subnetwork, even if it does not have a large number of direct neighbours in that subnetwork or a high closeness centrality with respect to the whole network.

The standard closeness centrality $c_v$ \cite{Freeman1979} can be obtained from the $c_v^j$ as
\begin{equation}
c_v = \frac{N - 1}{\sum_j N_j \left(c_v^j\right)^{-1}}. \label{eq:closeness}
\end{equation}

\paragraph{Cross-betweenness centrality}

For any vertex $w \in V$, cross-betweenness centrality $b_w^{ij}$ indicates its role for mediating or controlling interactions/communication between two subnetworks $G_i$ and $G_j$,
\begin{eqnarray}
b_w^{ij} &=& \sum_{p \in V_i, q \in V_j; p,q \neq w} \frac{\sigma_{pq}(w)}{\sigma_{pq}} \label{eq:cross_betweenness} \\ 
&=& b_w^{ji}, \nonumber
\end{eqnarray}
where $\sigma_{pq}$ gives the total number of shortest paths from $p$ to $q$ and $\sigma_{pq}(w)$ counts the number of shortest paths between $p$ and $q$ that include $w$ \cite{Freeman1979,Brandes2008}. While a vertex with high cross-degree (\emph{relational hub}) may function as a robust transmitter between two subnetworks, its functional redundancy for communication is evaluated by cross-between\-ness centrality. E.g., a relational hub with high cross-between\-ness is more vulnerable to attack or failure with respect to the interaction of two subnetworks than another one with low cross-betweenness and, hence, high redundancy.

The standard betweenness centrality $b_w$ \cite{Freeman1979} of the full network is obtained by summing up the contributions from all pairs of subnetworks,
\begin{equation}
b_w = \sum_{ij} b_w^{ij}. \label{eq:betweenness}
\end{equation}
In this sense, cross-betweenness centrality can be seen as a decomposition of the standard betweenness centrality with respect to a certain partition of the full network. Cross-betweenness is a generalisation of the measure $Q_2$ defined by Flom \textit{et al.} that is restricted to networks consisting of only two different sorts of vertices or two subnetworks in our terminology \cite{Brandes2008,Flom2004}.

\subsection{Global measures}
\label{sec:global_measures}

\paragraph{Cross-edge density}

$\rho_{ij}$ measures the density of connections between distinct subnetworks $G_i$ and $G_j$,
\begin{eqnarray}
\rho_{ij} &=& \frac{|E_{ij}|}{N_i N_j} \label{eq:cross_edge_density}\\
&=& \frac{\left<k_v^j\right>_{v \in V_i}}{N_j}\nonumber \\
&=& \rho_{ji}. \nonumber
\end{eqnarray}
Two subnetworks can be considered to be well separated topologically if their internal edge densities $\rho_i,\rho_j$ are clearly larger than their cross-edge density, i.e., $\rho_{ij} \ll \rho_i, \rho_j$. More generally, subnetworks form communities of the full network \cite{Fortunato2010} if this inequality holds for all pairs of subnetworks. In other words, in this situation the partition of the full network $G$ induced by the subnetworks $G_i$ would give rise to a high modularity \cite{Newman2004}. It should be stressed, however, that there exists a multitude of other definitions of communities \cite{Fortunato2010}. Our general framework does not require the chosen partition to be consistent with any such definition as long as it allows the resulting cross-network measures to be readily interpretable.

\paragraph{Global cross-clustering coefficient}

$\mathcal{C}_{ij}$ is an estimate of the probability of vertices from subnetwork $G_i$ to have mutually connected neighbours within subnetwork $G_j$,
\begin{eqnarray}
\mathcal{C}_{ij} &=& \left<\mathcal{C}_v^{ij}\right>_{v \in V_i} \label{eq:cross_clustering}\\
&=& \frac{1}{N_i} \sum_{v \in V_i, k_v^{j}>1} \frac{\sum_{p \neq q \in V_j} A_{vp} A_{pq} A_{qv}}{\sum_{p \neq q \in V_j} A_{vp}  A_{vq}} \nonumber.
\end{eqnarray}
It is important to note that in contrast to cross-edge density and cross-average path length the cross-clustering coefficient is not a symmetric property of two subnetworks, i.e., $\mathcal{C}_{ij} \neq \mathcal{C}_{ji}$. As was shown above, we expect $\mathcal{C}_{ij}=\mathcal{O}(\rho_j)$ if the interaction structure of subnetworks $G_i$ and $G_j$ is random, i.e., cross edges are distributed randomly and independently between the two subnetworks. In contrast, an organised interdependence would induce $\mathcal{C}_{ij}\gg \rho_j$ or $\mathcal{C}_{ij}\ll \rho_j$.

\paragraph{Cross-transitivity}

$\mathcal{T}_{ij}$ is the probability that two vertices in subnetwork $G_j$ are connected if they have a common neighbour in subnetwork $G_i$,
\begin{equation}
\mathcal{T}_{ij} = \frac{\sum_{v \in V_i; p \neq q \in V_j} A_{vp} A_{pq} A_{qv}}{\sum_{v \in V_i; p \neq q \in V_j} A_{vp}  A_{vq}}. \label{eq:cross_transitivity}
\end{equation}
Here also $\mathcal{T}_{ij} \neq \mathcal{T}_{ji}$. As the global cross-clustering coefficient $\mathcal{C}_{ij}$, $\mathcal{T}_{ij}$ is a measure of the degree of organisation of the subnetworks' topological interdependence. Analogous to the standard version of transitivity \cite{Newman2003}, $\mathcal{T}_{ij}$ tends to weigh contributions of vertices in $G_i$ with low cross-degree $k_v^j$ less heavily than $\mathcal{C}_{ij}$. From another point of view, in the average of Eq. (\ref{eq:cross_transitivity}) edges have equal weights, while Eq. (\ref{eq:cross_clustering}) has the same property with respect to vertices. This peculiarity has to be born in mind when interpreting the values of $\mathcal{T}_{ij}$ and $\mathcal{C}_{ij}$ for general complex networks of networks. Given a fully random interconnectivity structure, the expectation value for cross-transitivity is $\left<\mathcal{T}_{ij}\right> = \rho_{j}$.

\paragraph{Cross-average path length}

$\mathcal{L}_{ij}$ measures the topological closeness of two subnetworks $G_i$ and $G_j$,
\begin{eqnarray}
\mathcal{L}_{ij} &=& \frac{1}{N_i N_j - M_{ij}} \sum_{v \in V_i, q \in V_j} d_{vq} \label{eq:cross_average_path_length} \\
&=& \mathcal{L}_{ji}, \nonumber
\end{eqnarray}
where $M_{ij}$ is the number of pairs $(v \in V_i, q \in V_j)$ which are not mutually reachable. Hence, $\mathcal{L}_{ij}$ measures the average length of existing shortest paths between subnetworks $G_i$ and $G_j$. $\mathcal{L}_{ij}$ can be interpreted as a measure of the efficiency of interaction between two subnetworks. Subnetworks with low  $\mathcal{L}_{ij}$ are closely interwoven and are likely to show a high degree of functional interdependence, while those with high $\mathcal{L}_{ij}$ are topologically more separated and likely to be dynamically and functionally more independent of each other. If $M_{ij}=0$, $\mathcal{L}_{ij}$ is related to the cross-closeness centralities $c_v^{ij}$ via
\begin{equation}
\mathcal{L}_{ij} = \frac{1}{N_i} \sum_{v \in V_i} \left(c_v^{ij}\right)^{-1}. \label{eq:apl_closeness}
\end{equation}

%
%
\section{Application: Analysing the vertical dynamical structure of the Earth's atmosphere}
\label{sec:application}

\begin{figure}
\centering
\resizebox{1.0\columnwidth}{!}{%
  \includegraphics{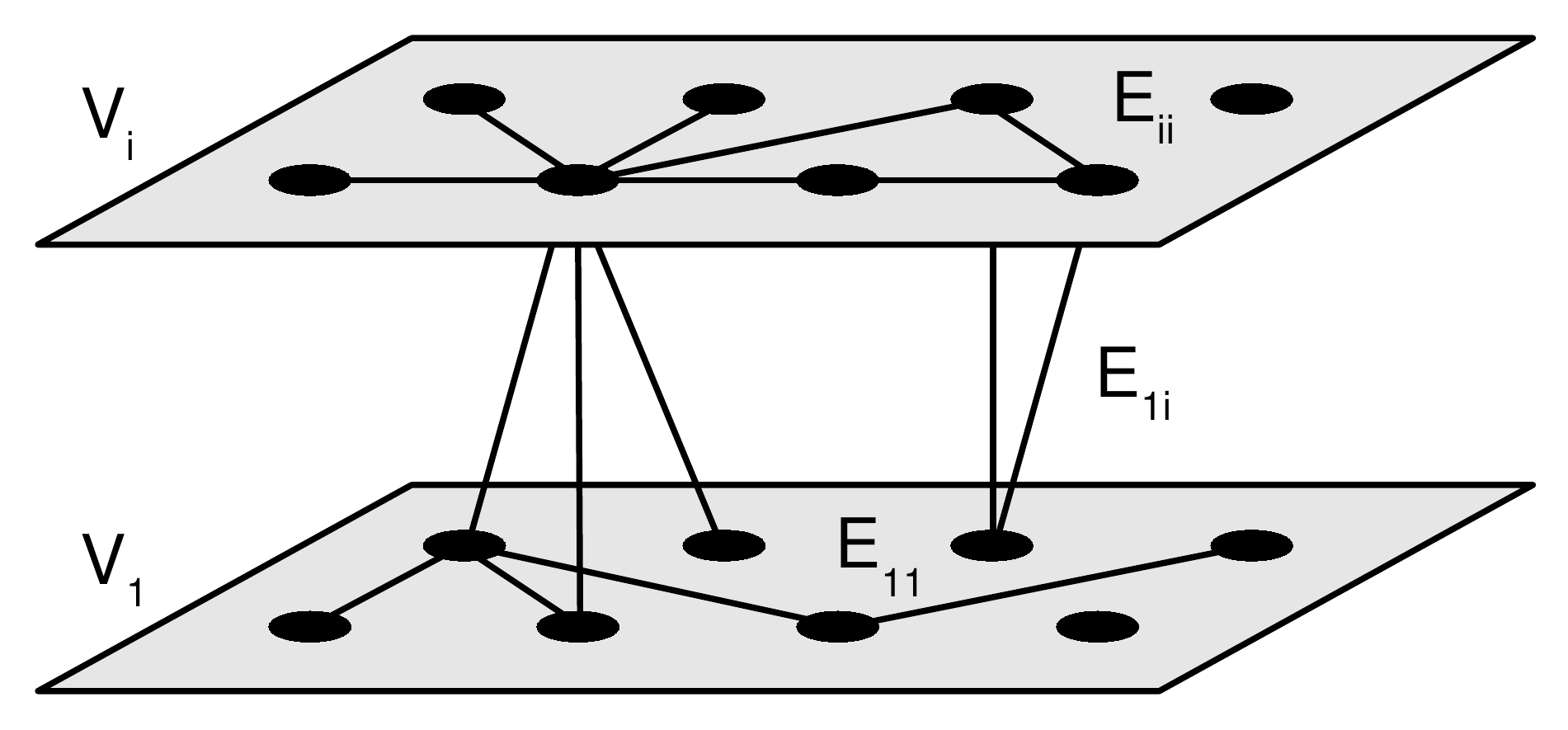}
}
\caption{Illustration of a coupled climate subnetwork as it is constructed in this work, where $V_1$ denotes the set of vertices in the near ground subnetwork and $V_i$ that of another isobaric surface higher up in the atmosphere. $E_{11}, E_{ii}$ are sets of internal edges of the two isobaric surfaces or subnetworks describing the statistical relationships within each isobaric surface, while $E_{1i}$ contains information on their mutual statistical interdependencies.}
\label{fig:coupled_network_sketch}
\end{figure}

Climatologists are interested in studying the correlation structure of climatological field variables in order to find spatial as well as temporal patterns accounting for a large fraction of the field's variance, commonly relying on established linear methods such as principal component analysis (termed empirical orthogonal functional analysis in climatological parlance) or singular spectrum analysis \cite{Kutzbach1967,Wallace1981,Vautard1989,Mudelsee2010}. Recently, complex climate networks have been successfully used to analyse single fields of climate observables combining methods from nonlinear time series analysis and network theory\cite{Donges2009,Donges2009a,Donner2008,Gozolchiani2008,Tsonis2004,Tsonis2008a,Tsonis2008b,Yamasaki2008}.

The formerly mentioned classical approach has been generalized to find coupled patterns in climate data investigating the correlation structure between fields of different climatological observables using techniques such as canonical correlation analysis or singular value decomposition of the covariance matrix \cite{Bretherton1992}. In analogy to the study of coupled patterns, here for the first time we expand the climate network approach to analyse the dynamical interrelationships between two different climatological fields by constructing \emph{coupled climate subnetworks} and investigating them within the interacting networks framework introduced above. In other words, the correlation structure within and between two sets of discrete field observables $X_i(t), Y_j(t)$ is mapped to a complex coupled climate subnetwork (Fig. \ref{fig:coupled_network_sketch}), where $i$ and $j$ are spatial indices and $t$ denotes time.

In order to justify treating a coupled climate subnetwork as a network of networks, an \textit{ab initio} physical separation of the climatological fields is necessary regarding processes responsible for internal coupling within a single field and those mediating interactions between both fields, as is elaborated in the following. The Earth's quasi-spherical shape and almost homogeneous mass distribution result in a hydrostatic equilibrium in first-order approximation, implying a stable isobaric quasi-horizontal stratification and therefore a strong buoyancy constraint \cite{Salby1996}. Local heating of the Earth's surface and atmosphere due to temporally and spatially varying solar radiation induces minor disturbances of the system, which give rise to weather variability and are propagated by advection, diffusion (dominantly turbulent in the homosphere) and convection processes. Convection processes lead to vertical movement resulting in pressure gradients which are balanced by quasi-horizontal geostrophic winds along isobares. An important mechanism is slant convection arising as an adjustment of baroclinic instabilities \cite{Pedlosky1979}.

Geopotential height $Z(\vartheta,\phi,h)$ is a vertical coordinate referenced to the Earth's mean sea level which takes into account the variation of gravitational acceleration $g(\vartheta,\phi,h)$ with latitude $\vartheta$, longitude $\phi$ and geometrical height $h$. It is defined as 
\begin{equation}
Z(\vartheta,\phi,h) = \frac{1}{g_0} \int_0^h g(\vartheta,\phi,z) dz,
\end{equation}
where $g_0$ denotes the standard gravitational acceleration at mean sea level \cite{Salby1996}. $Z(\vartheta,\phi,h)$ is approximately equivalent to geometrical height $h$ within the homosphere, i.e., the lower portion of the atmosphere we consider in this work. The geopotential height $Z(\vartheta,\phi,P)$ of a certain pressure level $P$ is defined as the geopotential height necessary to reach the given pressure $P$. In meteorology and climatology, the field $Z(\vartheta,\phi,P)$ is frequently used as an equivalent and convenient representation of the three-dimensional atmospheric pressure field $P(\vartheta,\phi,Z)$. Therefore the discretised and vertically resolved geopotential height field $Z_v^i(t)$ sampled at predefined points $v$ on isobaric surfaces $i$ at pressure $P_i$ captures the dynamics of both the geostrophic wind field as well as convection processes and, hence, reflects global weather and climate dynamics to a good approximation \cite{Salby1996}. Given the distinct physical processes behind vertical and quasi-horizontal atmospheric dynamics described above, it is hence feasible to apply the interacting networks approach, treating the induced subgraphs of vertices lying on the same isobaric surface $i$ as distinct subnetworks $G_i$. In this paper we specifically focus on the interaction structure between near ground and upper level atmospheric dynamics, which is particularly interesting as a large portion of the solar forcing driving atmospheric dynamics takes place on the Earth's surface (Fig. \ref{fig:coupled_network_sketch}).

To better understand our coupled climate subnetwork approach, one should be aware of the strong analogy existing between the concepts of climate networks and functional brain networks studied in neuroscience and medicine \cite{Zamora2009,Zamora2010,Zhou2006,Zhou2007,Bialonski2010}. Both types of networks describe statistical similarity relationships between spatially embedded time series using the same methods of time series analysis, but relying on distinct data sources, i.e., the climate system and the mammalian brain. Now there exist two fundamentally different types of networks: (i)~\emph{Structural} (or \emph{anatomical}) networks, on the one hand, reflect the topological structure of existing ties between objects (e.g., computers, neurons, columns of neurons), referring to either physical connections (e.g., internet, power grids, neuronal networks) or abstract relations (e.g., world wide web, social networks, citation networks). (ii)~\emph{Functional networks}, on the other hand, including functional brain networks and complex climate networks, are extracted from an underlying system by detecting and assessing similarities in the dynamical behaviour of its components. In other words, structural networks represent \textit{a priori} knowledge on a system's internal structure on a certain level of abstraction, whereas functional networks are inferred solely from the measured or simulated dynamics of subsystems, usually without including any additional information. Hence, in contrast to structural networks, the resulting topological interconnections in functional networks do not directly allow to draw conclusions on a causal interrelationship between vertices. When constructed and compared for the same system, e.g. a neuronal network, both types of networks will usually not be identical. This implies that special emphasis has to be put on physical arguments when interpreting topological features of climate networks. A further noteworthy duality exists between spatial similarity networks based on fields of time series such as the climate networks discussed above, and temporal similarity networks, e.g., recurrence networks \cite{Donner2010a,Donner2010b,Marwan2009}, representing a single time series.

\subsection{Data}
\label{sec:data}

\begin{table}[b]
\caption{Pressure $P_i$ and associated mean geopotential height $Z_i$ (Eq. (\ref{eq:mean_geopotential_height})) for each isobaric surface $i$ in the NCEP/NCAR Reanalysis 1 reconstruction of the geopotential height field.}
\label{tab:pressure_height}
\centering
\begin{tabular}{c|c|c}
\hline\noalign{\smallskip}
$i$ & $P_i$ [mbar] & $Z_i$ [km]  \\
\noalign{\smallskip}\hline\noalign{\smallskip}
1 &  1000 &      0.1\\
2 &   925  &     0.8\\
3 &   850  &     1.5\\
4 &   700  &     3.0\\
5 &   600  &     4.3\\
6 &   500  &     5.7\\
7 &   400  &     7.3\\
8 &   300  &     9.3\\
9 &   250  &    10.6\\
10 &   200  &    12.0\\
11 &   150  &    13.8\\
12 &  100  &   16.3\\
13 &    70   &   18.5\\
14 &    50   &  20.5\\
15 &    30   &   23.8\\
16 &    20   &   26.4\\
17 &    10   &   30.9\\
\noalign{\smallskip}\hline
\end{tabular}
\end{table}

To construct coupled climate subnetworks capturing longer-term dynamics of the geostrophic wind field as well as large-scale convection processes, we rely on the global month\-ly averaged and vertically resolved atmospheric geopotential height field covering the troposphere and the lower stratosphere. We use Reanalysis 1 data provided by the National Center for Environmental Prediction/National Center for Atmospheric Research (NCEP/\-NCAR) \cite{Kistler2001}. For each of the $17$ isobaric surfaces $P_i$ the NCEP/NCAR data is given on an equally spaced spherical grid with a latitudinal and longitudinal resolution of $2.5^\circ \times 2.5^\circ$, resulting in $10,266$ grid points. Using this type of grid for network construction would induce biases in the statistical properties of climate networks, since the area covered by each grid point is not uniform but decreases towards the poles like the cosine of latitude \cite{Tsonis2008b,Heitzig2010}. To avoid these effects, we choose to project the data to an icosahedral grid \cite{Heikes1995} of $N_i=2,562$ time series $Z_v^i(t)$ with $v \in \{1,\dots, N_i \}$ for each isobaric surface $i$, respectively, at pressure $P_i$ using the conservative interpolation scheme described in \cite{Jones1998}. Each isobaric surface $i$ may be associated to an average geopotential height 
\begin{equation}
Z_i=\left<Z_v^i(t)\right>_{v,t} \label{eq:mean_geopotential_height}
\end{equation}
by averaging over time $t$ and all grid points $v$ contained within this level (Tab. \ref{tab:pressure_height}). The time series $Z_v^i(t)$ contain $734$ monthly averaged data points from January 1948 until February 2009. 

Relying on the icosahedral grid, which is used by the German Weather Service for its operational global weather forecast model \textit{Global Modell Extended} (GME) and referred to as ``triangular grid" \cite{Majewski2002}, guarantees nearly uniform grid cell areas within each isobaric surface $i$. The differences in grid cell area between different isobaric surfaces due to their varying distance from the Earth's surface are negligible since the maximum vertical separation of isobaric surfaces ($\approx 30$\,km, see Tab. \ref{tab:pressure_height}) is much smaller than the Earth's mean radius $a \approx 6,370$\,km.

As a final step of preprocessing the data, we calculate climatological anomaly time series $\hat{Z}_v^i(t)$ by phase averaging (see \cite{Donges2009a}) to remove the leading order effect of the annual cycle from the geopotential height time series $Z_v^i(t)$. This helps to avoid spurious correlations solely due to the solar forcing common to all time series.

\subsection{Network construction}
\label{sec:network_construction}

We now construct a sequence of pairwise coupled climate subnetworks based on statistical interrelationships within the three-dimensional geopotential height ano\-maly field by using the linear Pearson correlation at zero lag. As was shown in \cite{Donges2009a}, linear correlation measures are sufficient for a first overview study like the present work, because they capture the great majority of statistical interrelationships within fields of smooth (non-intermittent) climatological variables like temperature or geopotential height. The so obtained networks describe both the intrinsic structure of a single isobaric surface as well as the interaction structure between two isobaric surfaces, in other words comprising horizontal as well as vertical interdependencies of the spatially embedded time series. First the $N_i, N_j=2,562$ time series of two isobaric surfaces $i,j$ are relabeled using indices $p,q\in\{1,\dots,N\}$, $N = N_i + N_j$, $\hat{Z}_v^i(t), \hat{Z}_v^j(t)\rightarrow \hat{Z}_p(t), \hat{Z}_q(t)$, where the fully coupled climate subnetworks contain $N=5,124$ vertices. The anomaly time series $\hat{Z}_{p,q}(t)$ are identified with vertices $p,q$ of the coupled climate subnetwork and subsequently collected in sets $V_i, V_j$ based on their native isobaric surface. The classic Pearson correlation between any pair of $\left\lbrace \hat{Z}_p(t), \hat{Z}_q(t) \right\rbrace$ is then given by
\begin{equation}
P_{pq}^{i,j} = \frac{\left<\left(\hat{Z}_p(t)-\mu_p\right) \left(\hat{Z}_q(t)-\mu_q\right)\right>_t}{\sigma_p \sigma_q},
\end{equation}
where $\mu_{p,q}$ and $\sigma_{p,q}$ are the mean and standard deviation of the $\hat{Z}_{p,q}(t)$, respectively. Edges $\{p,q\}$ are introduced into the coupled climate subnetwork if the absolute value of Pearson correlation $|P_{pq}^{i,j}|$ at zero lag between both time series  $\hat{Z}_p(t)$ and  $\hat{Z}_q(t)$ exceeds a certain threshold value $0\leq T \leq1$. The threshold should be carefully chosen to ensure that only statistically significant and reasonably strong correlations are included in the network \cite{Donges2009a}. Because an optimal value of $T$ is neither easy to define nor to determine (it would even be possible to determine a threshold $T_{pq}$ for each pair of time series $p,q$ separately), we will present results for various thresholds below (Sect.~\ref{sec:results}). Furthermore it would be feasible to avoid the choice of a threshold altogether by including a function of the correlation measure $w_{pq}^{i,j}=f(P_{pq}^{i,j})$ into the network analysis as edge weights $w_{pq}^{i,j}$. However, we do not follow this research avenue here as it introduces as a new complication the choice of a meaningful transfer function $f$ for mapping correlations to edge weights depending on the interpretation of a specific network measure. In contrast, this work focusses on the topology of statistical interrelationships between different isobaric surfaces on the network level. The consideration of edge weights will be an interesting subject of future work.

Finally, the coupled climate subnetworks are completely described by the adjacency matrices
\begin{equation}
A_{pq}^{i,j} = \Theta \left(\left|P_{pq}^{i,j}\right| - T\right) - \delta_{pq},
\end{equation}
where $\Theta(\cdot)$ is the Heaviside function and Kronecker's delta $\delta_{pq}$ indicates that artificial self-loops are not considered. The subnetworks $G_i,G_j$ representing the correlation structure within each isobaric surface $i,j$ are the induced subgraphs of the sets $V_i,V_j$ embedded within the coupled climate subnetwork $G^{i,j}$ described by the adjacency matrix $A_{pq}^{i,j}$. In the following, local as well as global (cross-)network measures $m_v^{ij}, m_{ij}$ will be calculated from the coupled climate subnetwork $G^{i,j}$ consisting of two isobaric subnetworks $G_i$ and $G_j$ and their interaction structure $E_{ij}$, i.e., $G^{i,j} = (V_i \cup V_j, E_{ii} \cup E_{jj} \cup E_{ij})$.

\subsection{Results}
\label{sec:results}

\paragraph{Global measures}

%
%
\begin{figure*}[t!]
\centering
\resizebox{1.0\columnwidth}{!}{%
  \includegraphics{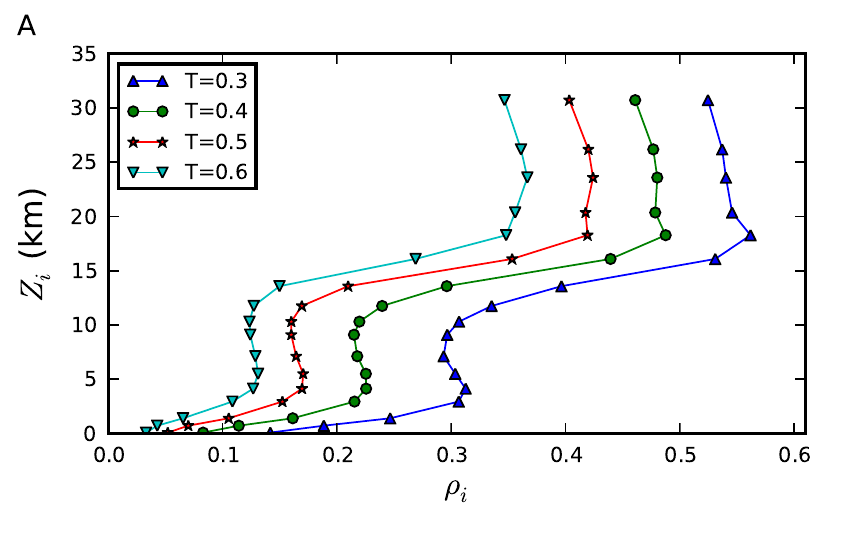}
}
\resizebox{1.0\columnwidth}{!}{%
  \includegraphics{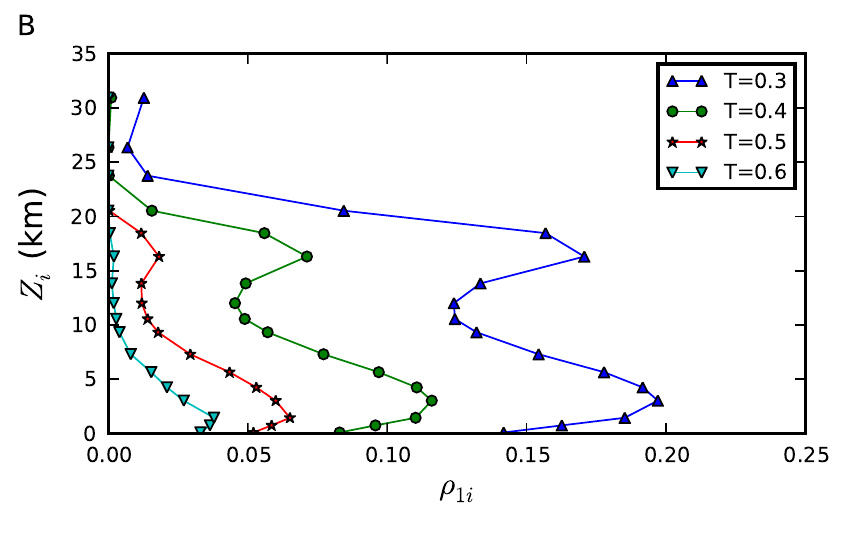}
}
\resizebox{1.0\columnwidth}{!}{%
  \includegraphics{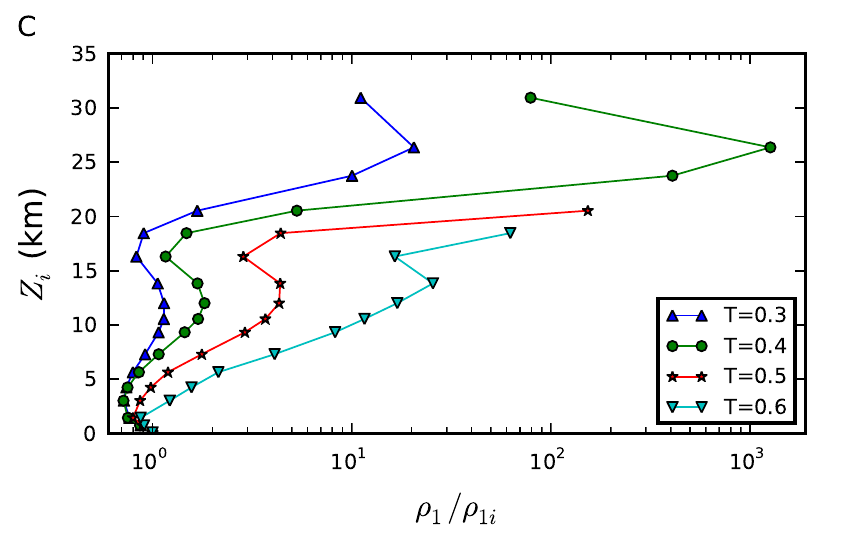}
}
\resizebox{1.0\columnwidth}{!}{%
  \includegraphics{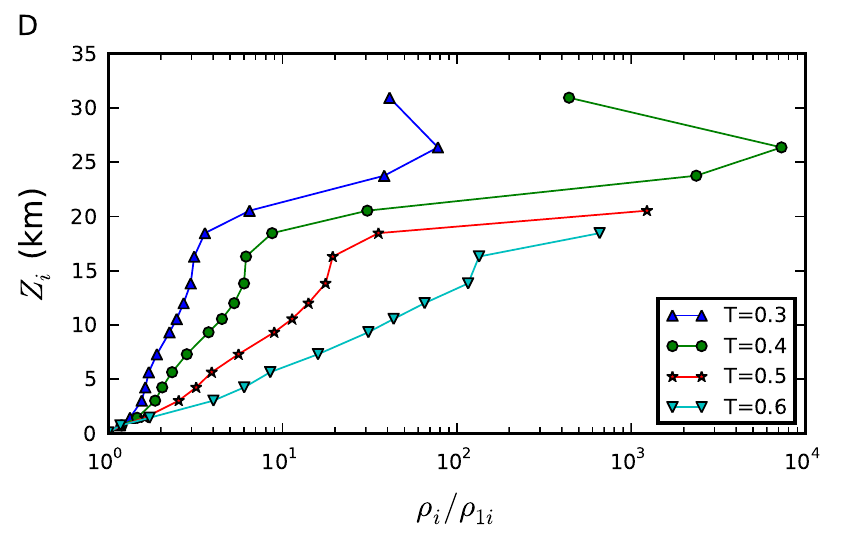}
}
\caption{(A) Internal edge density $\rho_{i}$ of all subnetworks on isobaric surfaces $i$, (B) cross-edge density $\rho_{1i}$ between the surface level $1$ and all other isobaric surfaces $i$ of average height $Z_i$, (C) the ratio $\rho_1/\rho_{1i}$ and (D) the ratio $\rho_i/\rho_{1i}$ in the coupled geopotential height climate network for various thresholds $T$. Note the inversions of cross-edge density at $Z \approx 1-3, 12, 16$ and $26$\,km becoming increasingly pronounced for decreasing threshold $T$.}
\label{fig:link_densities}
\end{figure*}

The available data describes the dynamics of geopotential height in the lower homosphere, encompassing the troposphere and lower stratosphere, where most atmospheric dynamical processes relevant for the Earth's climate system are concentrated to \cite{Salby1996}. In the following we will investigate which aspects of atmospheric dynamics can be revealed by analysing the sequence of coupled climate subnetworks $G^{1,i}, i=1,...,17$. 

It is pivotal to be aware that for similarity networks constructed from spatio-temporal data, such as the coupled climate subnetworks studied here, network structure is subject to statistical uncertainties \cite{Kramer2009}. One consequence is that average path length and global clustering coefficient cannot be considered as useful order parameters for network classification \cite{Bialonski2010}. Both are biased by spurious or missing links due to the network construction algorithm and local correlations between spatially close observations. Particularly, this implies that it is not meaningful to classify interaction networks as small-world networks \cite{Watts1998}, Erd\H{o}s-R\'enyi type random networks or grid-like regular networks, as is common practice for other types of networks \cite{Newman2003,Boccaletti2006}. Since the corresponding arguments of Bialonski \textit{et al.} \cite{Bialonski2010} also hold for the modified clustering coefficients, transitivities and average path lengths defined in Sect.~\ref{sec:theory}, we in the following do not consider the absolute values of these measures. In contrast, here we are solely interested in their relative changes conveying information on the varying interaction structure within and between different isobaric surfaces of the atmosphere.

Our first and fundamental observation is that the cross-edge density $\rho_{1i}$ (Fig.~\ref{fig:link_densities}B) between the near-surface and all other isobaric surfaces $i$ is always smaller than and well separated from the internal edge density of the upper isobaric subnetwork $\rho_{i}$ (Fig.~\ref{fig:link_densities}A) for all considered thresholds $T$. Moreover, the ratio of the edge densities $\rho_{i}/\rho_{1i}$ and, hence, the physical separation of the underlying dynamics is increasing with height $Z_i$ (Fig.~\ref{fig:link_densities}D). Approximately the same holds for the internal edge density $\rho_{1}$ of the near-surface subnetwork which is considerably larger than $\rho_{1i}$ for most $Z_i$ and $T$ (Fig.~\ref{fig:link_densities}C). This observation reflects topologically the dynamical separation of atmospheric processes within and between isobaric surfaces $i$, respectively, that led to identifying them with subnetworks in the first place (see the introduction to Sec. \ref{sec:application}). Nevertheless, $\rho_{1}$ can be slightly smaller than the cross-edge density $\rho_{1i}$, particularly close to the two pronounced minima of the ratio $\rho_{1}/\rho_{1i}$ (the overall minimum value is $\min_{i,T} \rho_{1}/\rho_{1i} \approx 0.7$, see Fig.~\ref{fig:link_densities}C). This finding, however, does not challenge the applicability of our method as our framework does not require that subnetworks constitute communities of the full network (Sec. \ref{sec:global_measures}).

The cross-edge density $\rho_{1i}$ displays prominent extrema with varying height $Z_i$ of the isobaric surface $i$, which become more pronounced for decreasing threshold $T$ (Fig.~\ref{fig:link_densities}B). Two maxima of $\rho_{1i}$ are located at $1-3$\,km and $16$\,km, while a minimum is found at $12$\,km. A much more weakly developed inversion of cross-edge density $\rho_{1i}$ occurs at $26$\,km, but it is only visible for small $T$. These findings indicate that correlations of large-scale quasi-geostrophic wind dynamics are significantly increased between the near-ground and higher isobaric surfaces at approx. $1-3$\,km and $16$\,km. The superficially similar inversions in internal edge density $\rho_{i}$ with two maxima at $4-6$\,km, $18-24$\,km and a minimum at $7-11$\,km have to be carefully distinguished from those of cross-edge density $\rho_{1i}$ (Fig.~\ref{fig:link_densities}A). First, the geopotential height intervals within which the respective extrema are observed for different $T$ do not overlap. Second, on the one hand, recall that in contrast to $\rho_{1i}$ the internal edge density $\rho_{i}$ measures dynamical correlations occurring within a quasi-horizontal isobaric surface at pressure $P_i$. On the other hand, the physical processes acting within isobaric surfaces (geostrophic wind, planetary Rossby waves, gravity waves) and between them (convection, turbulent mixing), which are relevant for large scale dynamical coupling are distinctively different.

%
%
\begin{figure}
\centering
\resizebox{1.0\columnwidth}{!}{%
  \includegraphics{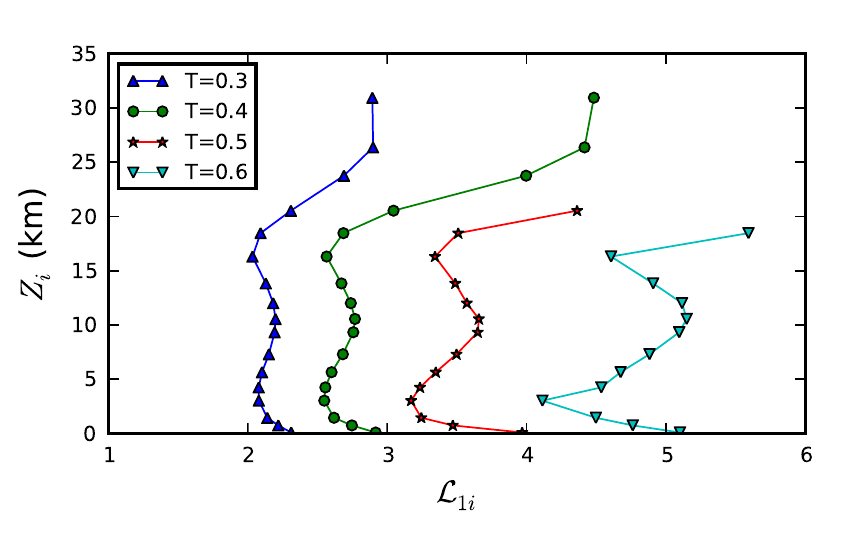}
}
\caption{The cross-average path length $\mathcal{L}_{1i}$ between the surface level $1$ and all other isobaric surfaces $i$ of average height $Z_i$ in the coupled geopotential height climate network for different thresholds $T$. In the lower stratosphere and for larger thresholds, $\mathcal{L}_{1i}$ is not defined due to vanishing cross-edge density $\rho_{1i}$ and, hence, the corresponding data points are not shown. $\mathcal{L}_{1i}$ shows inversions at similar height levels as the cross-edge density $\rho_{1i}$ (Fig. \ref{fig:link_densities}), which in contrast decrease in sharpness for deceasing threshold $T$.}
\label{fig:cross_apl}
\end{figure}

The cross-average path length $\mathcal{L}_{1i}$ between the near ground layer and all other isobaric surfaces possesses two minima at 3\,km and 16\,km as well as one maximum at 11\,km (Fig.~\ref{fig:cross_apl}). In contrast to cross-edge density, small values of $\mathcal{L}_{1i}$ imply tight dynamical relationships between two isobaric surfaces, while large values indicate a weaker coupling. Hence, cross-average path length consistently behaves complementarily to cross-edge density $\rho_{1i}$ and internal edge-density $\rho_{i}$, as shortest paths between two different subnetworks generally contain edges from both subnetworks implying their average length to decrease with increasing $\rho_{1i}$ and $\rho_{i}$. Interestingly, in contrast to the behaviour of $\rho_{1i}$ and $\rho_{i}$, the extrema in $\mathcal{L}_{1i}$ are rendered increasingly pronounced for increasing threshold $T$. Cross-average path length remains sensitive to variations in the topological closeness of two isobaric surfaces even as more and more edges $(p,q)$ of weaker correlation strength $|P_{pq}|$ are removed from the coupled climate subnetworks.

%
%
\begin{figure*}
\centering
\resizebox{1.0\textwidth}{!}{%
  \includegraphics{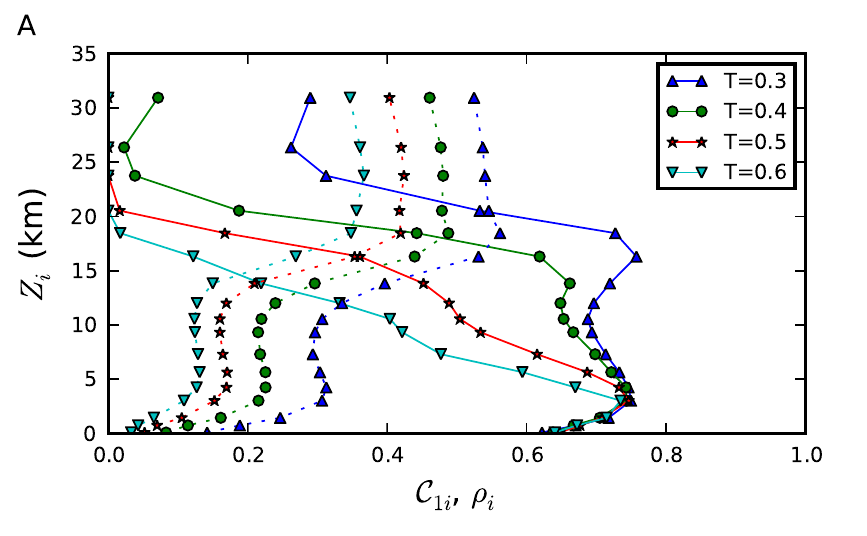}
  \includegraphics{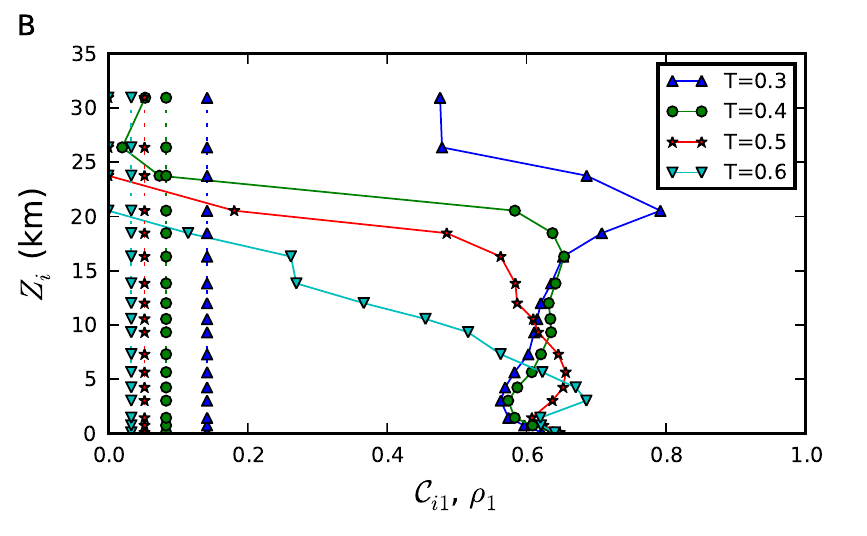}
}
\resizebox{1.0\textwidth}{!}{%
  \includegraphics{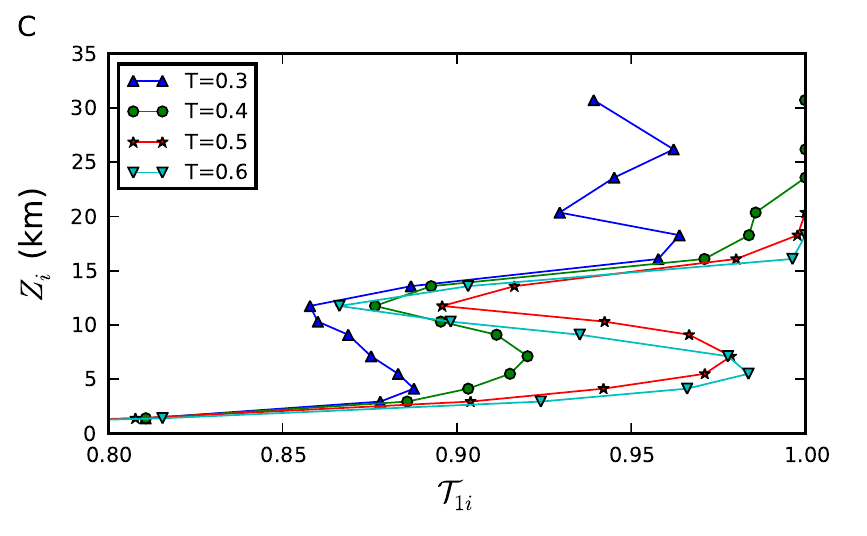}
  \includegraphics{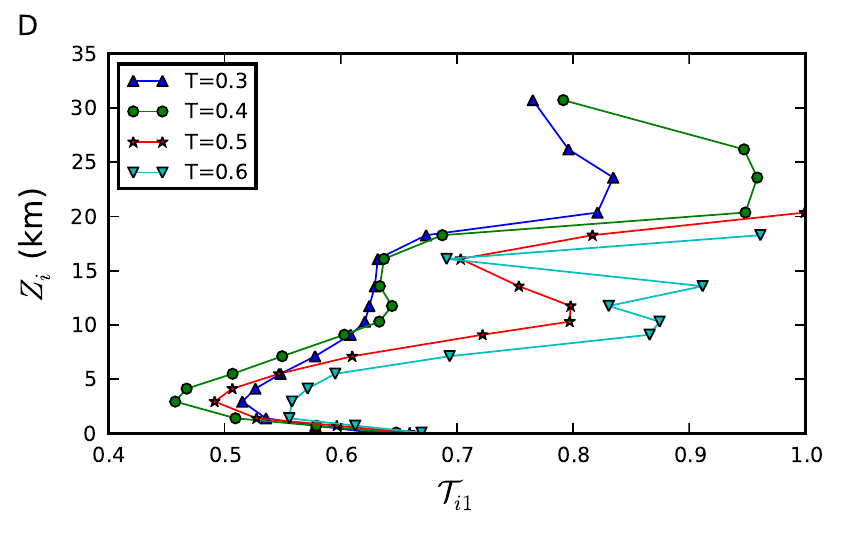}
}
\caption{Global cross-clustering coefficients (A) $\mathcal{C}_{1i}$ calculated ``upward" from the surface level $1$ to all other isobaric surfaces $i$ and (B) $\mathcal{C}_{i1}$ defined ``downward" (continuous lines), as well as cross-transitivities (C) $\mathcal{T}_{1i}$ (``upward") and (D) $\mathcal{T}_{i1}$ (``downward"). For comparison with the global cross-clustering coefficients $\left<\mathcal{C}_{ij}\right>$ expected for a fully random connectivity between isobaric surfaces, (A) and (B) also feature the expectation values $\left<\mathcal{C}_{1i}\right>=\rho_{i}$ and $\left<\mathcal{C}_{i1}\right>=\rho_{1}$, respectively (dashed lines).}
\label{fig:cross_clustering}
\end{figure*}

While the cross-network measures discussed so far are symmetric with respect to exchanging the involved subnetworks (see Sect.~\ref{sec:theory}), the two clustering measures global cross-clustering coefficient $\mathcal{C}_{ij}$ and cross-transitivity $\mathcal{T}_{ij}$ are intrinsically directional. Hence, as for the horizontally resolved cross-measures to be treated below, we are able to distinguish $\mathcal{C}_{1i}$ and $\mathcal{T}_{1i}$ pointing ``upward" from the near ground to higher isobaric surfaces from their counterparts $\mathcal{C}_{i1}$ and $\mathcal{T}_{i1}$ projecting ``downward". Both $\mathcal{C}_{1i}$ and $\mathcal{T}_{1i}$ consistently uncover that the probability of vertices within the near ground isobaric surface to have connected neighbours in higher isobaric surfaces reaches local maxima between $3-6$\,km and $14-16$\,km, whereas a local minimum is assumed at $11$\,km (Figs.~\ref{fig:cross_clustering}A and C). For both measures, all three inversions are only observed for low thresholds $T$. It is particularly interesting to compare the measured values of $\mathcal{C}_{1i}$ and $\mathcal{T}_{1i}$ with those expected for a fully random connectivity structure between isobaric surfaces (see Sect.~\ref{sec:global_measures}). The ``upward" pointing global cross-clustering coefficients $\mathcal{C}_{1i}$ are considerably larger than the expectations $\rho_i$  (indicated by dashed lines in Fig.~\ref{fig:cross_clustering}A) below $15$\,km, and markedly smaller for average geopotential heights above $20$\,km. Similarly, the ``upward" projecting cross-transitivity $\mathcal{T}_{1i}$ is significantly larger than the expected values $\rho_i$ for the fully random null model for all height levels (Fig.~\ref{fig:cross_clustering}C). 

Compared to their counterparts the behaviour of the ``downward" projecting measures $\mathcal{C}_{i1}$ and $\mathcal{T}_{i1}$ is less consistent. The global cross-clustering coefficient $\mathcal{C}_{i1}$ possesses two extrema at $3-6$\,km and $16-21$\,km, whereas the cross-transitivity $\mathcal{T}_{i1}$ takes local minima between $2-3$\,km and $14-16$\,km and local maxima between $12-14$\,km and at $24$\,km of geopotential height (Figs.~\ref{fig:cross_clustering}B and D). Note that ``downward" pointing cross-transitivity $\mathcal{T}_{i1}$ behaves complementarily to its ``upward" projecting counterpart $\mathcal{T}_{1i}$. Both clustering measures are significantly lar\-ger than their values $\rho_1$ expected for a fully random connectivity structure for nearly all height levels, only above $20$\,km the global cross-clustering coefficient $\mathcal{C}_{i1}$ is consistent with the expectation value for larger thresholds $T$. 

In summary, the clustering measures exhibit that the interaction topology between the near ground and higher isobaric surfaces is not consistent with a fully random null model, except for the lower stratosphere above $20$\,km. The same holds when comparing the observed values of the clustering measures to those expected from a more sophisticated null-model with fixed cross-degree sequences but otherwise random interaction topology (results not shown here). These findings highlight that the coupled climate subnetworks considered here have a nontrivial interaction topology, which is consistent with known features of the atmosphere's vertical dynamical structure and stratification and may contain additional information on previously unknown features of the atmosphere's general circulation \cite{Schneider2006,Hartmann2007}
 (see Sect.~\ref{sec:climatological_interpretation}). The large values of clustering measures observed for most height levels can be partly explained by spatial correlations between closely neighboured time series stemming from the continuity of the geopotential height field in conjunction with the intrinsic transitivity of the Pearson correlation coefficient $P_{pq}$ used for network construction \cite{Langford2001,Bialonski2010}. Directly comparing $\mathcal{C}_{i1}$ and $\mathcal{T}_{i1}$ as well as $\mathcal{C}_{1i}$ and $\mathcal{T}_{1i}$, respectively, furthermore clearly reveals the bias in the global cross-clustering coefficient, which leads to generally lower values for increasing height through weighing more strongly the contribution of the increasing number of vertices of low cross-degree induced by the decreasing trend in cross-edge density $\rho_{1i}$ (see Sect.~\ref{sec:global_measures}). Consequently, as for the standard versions of global clustering and transitivity \cite{Newman2003}, special care has to be taken when interpreting absolute values of global cross-clustering and cross-transitivity. The suggested best practice is to always consider the two measures simultaneously and to draw conclusions only from qualitative features exhibited by both of them.

%
%
\begin{figure*}
\centering
\resizebox{1.0\textwidth}{!}{%
  \includegraphics{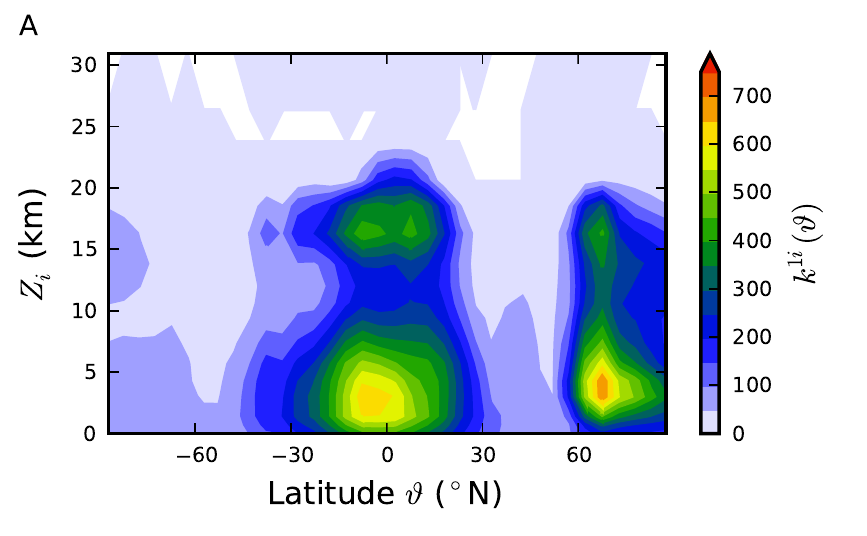}
  \includegraphics{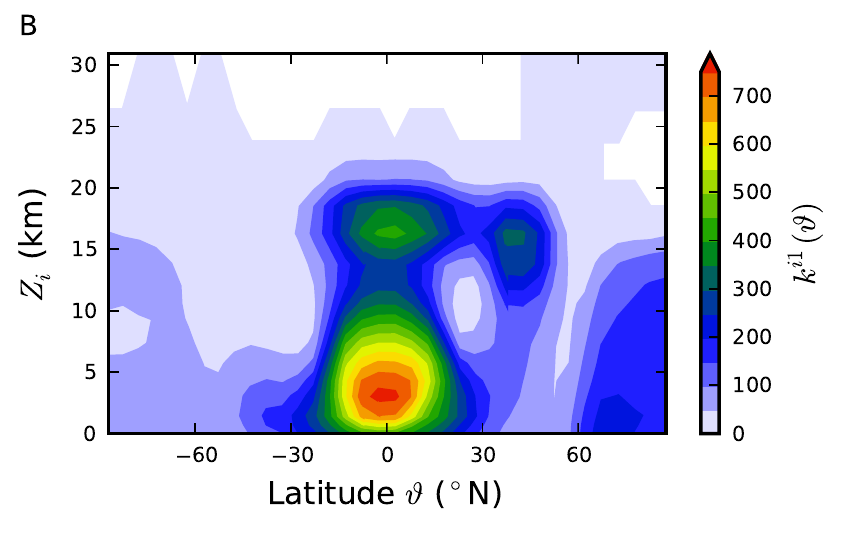}
}
\resizebox{1.0\textwidth}{!}{%
  \includegraphics{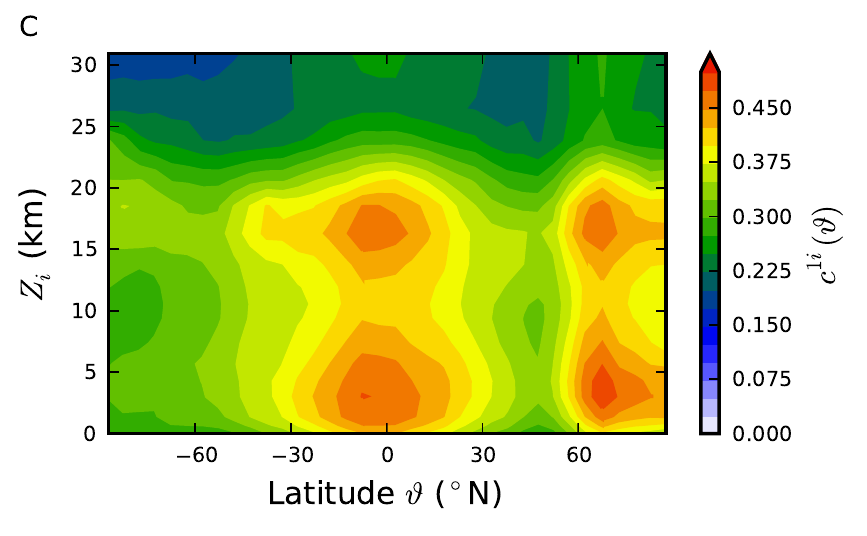}
  \includegraphics{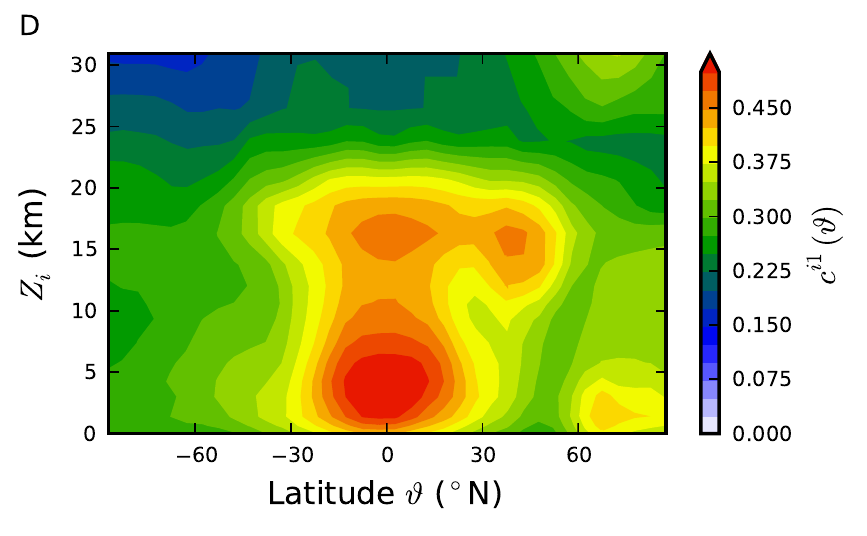}
}
\resizebox{1.0\textwidth}{!}{%
  \includegraphics{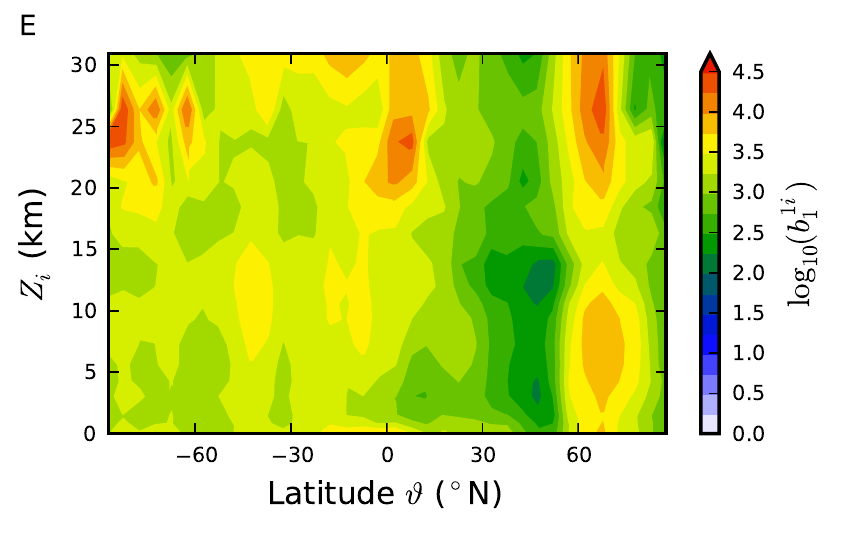}
  \includegraphics{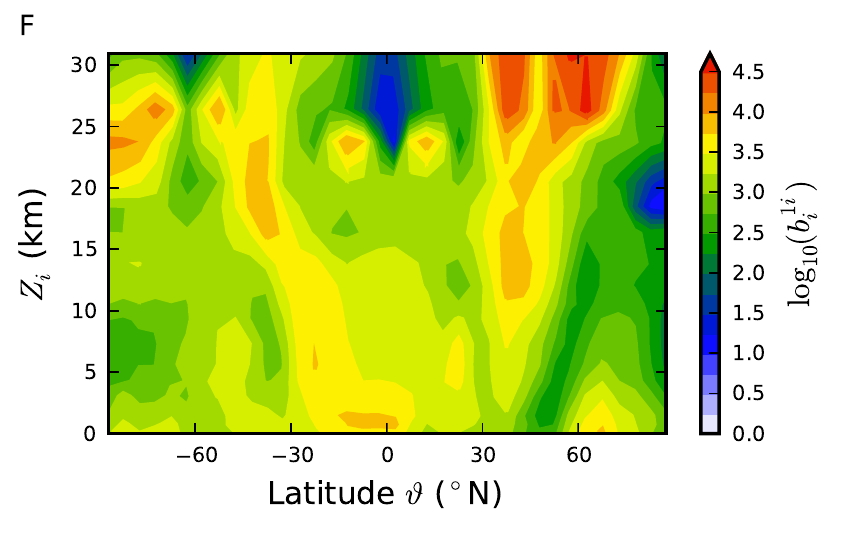}
}
\caption{Zonally averaged cross-degree centralities (A) $k^{1i}(\vartheta)$ pointing ``upward" from the near ground level $1$ to all other isobaric surfaces $i$ and (B) $k^{i1}(\vartheta)$ projecting ``downward", zonally averaged cross-closeness centralities (C) $c^{1i}(\vartheta)$ pointing ``upward" and (D) $c^{i1}(\vartheta)$ projecting ``downward", (E) $b_1^{1i}(\vartheta)$ near ground and (F) $b_i^{1i}(\vartheta)$ upper level component of zonally averaged cross-betweenness centrality for a threshold of $T=0.4$. Panel (B) can be interpreted to show the number of cross-edges connecting a certain volume element with the whole near ground isobaric surface, averaged along bands of approximately equal latitude (approximately because of the geodesic grid).}
\label{fig:cross_degree_closeness_vertical}
\end{figure*}

\begin{figure*}
\centering
\resizebox{1.0\textwidth}{!}{%
  \includegraphics{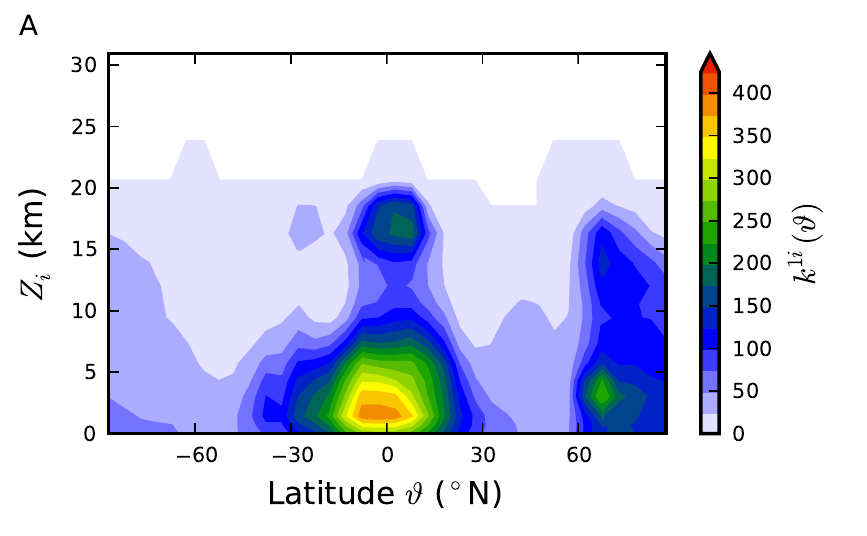}
  \includegraphics{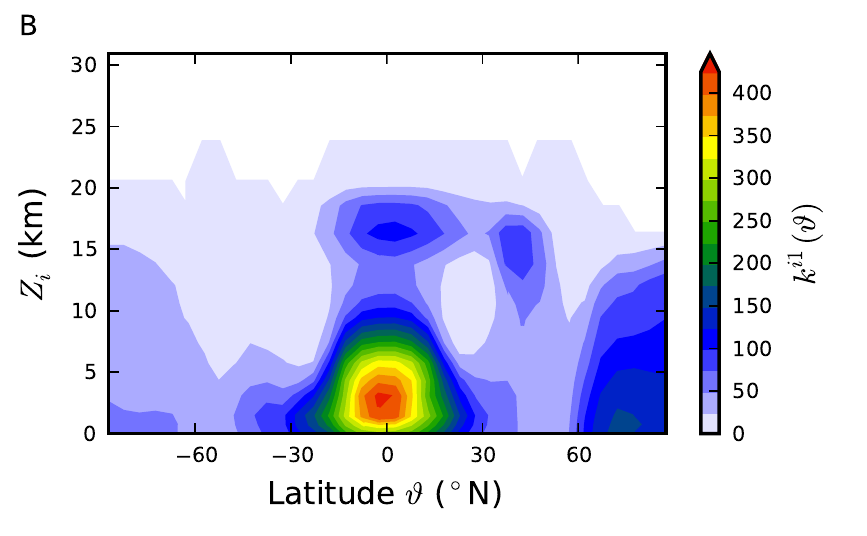}
}
\resizebox{1.0\textwidth}{!}{%
  \includegraphics{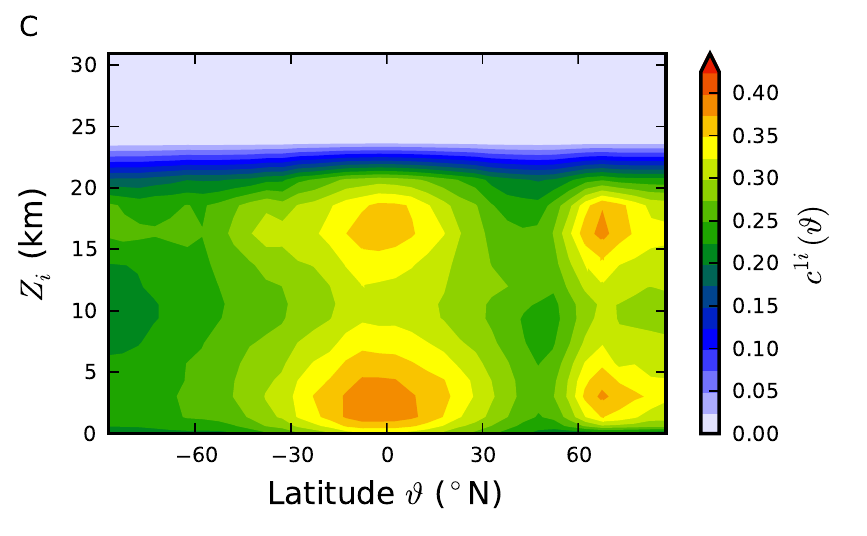}
  \includegraphics{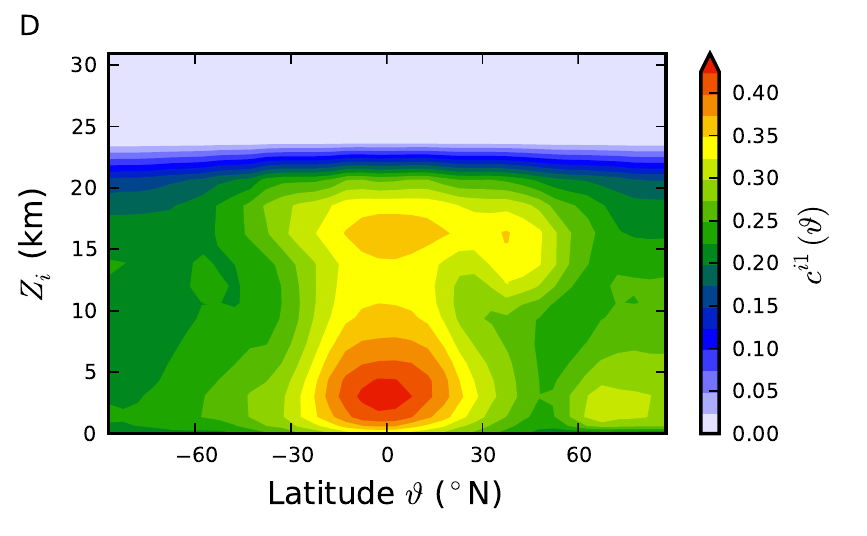}
}
\resizebox{1.0\textwidth}{!}{%
  \includegraphics{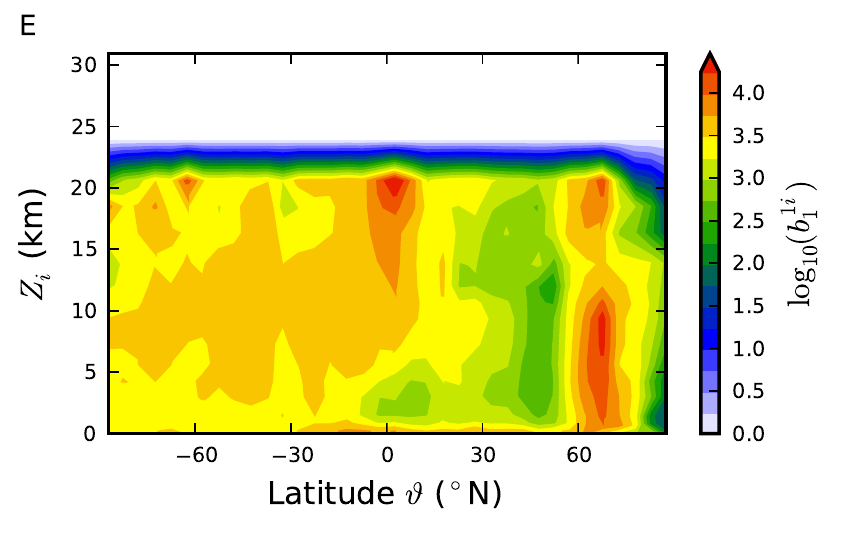}
  \includegraphics{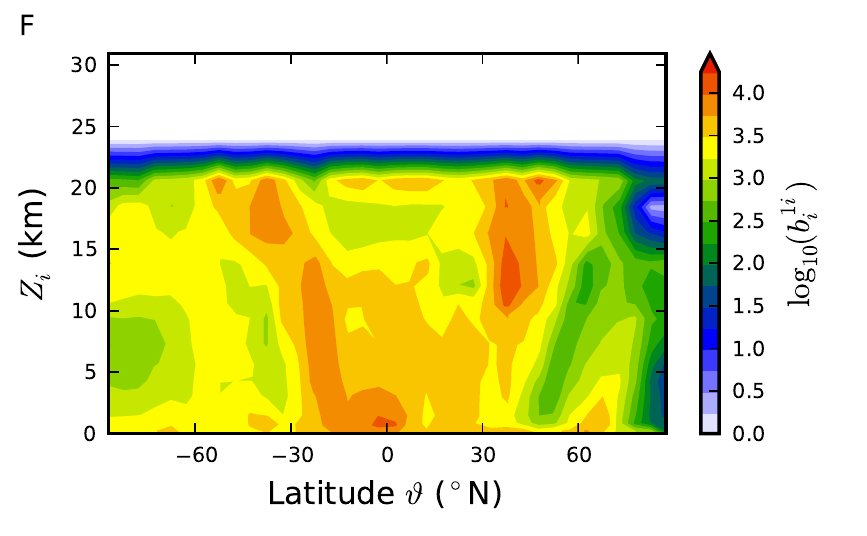}
}
\caption{As Fig. \ref{fig:cross_degree_closeness_vertical} with threshold $T=0.5$.}
\label{fig:cross_degree_closeness_vertical_T_0_5}
\end{figure*}

\begin{figure*}
\centering
\resizebox{1.0\textwidth}{!}{%
  \includegraphics{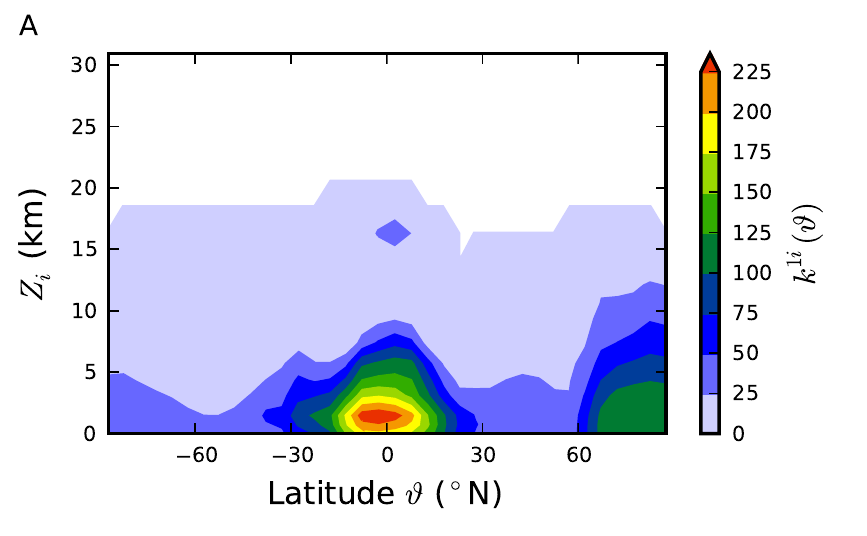}
  \includegraphics{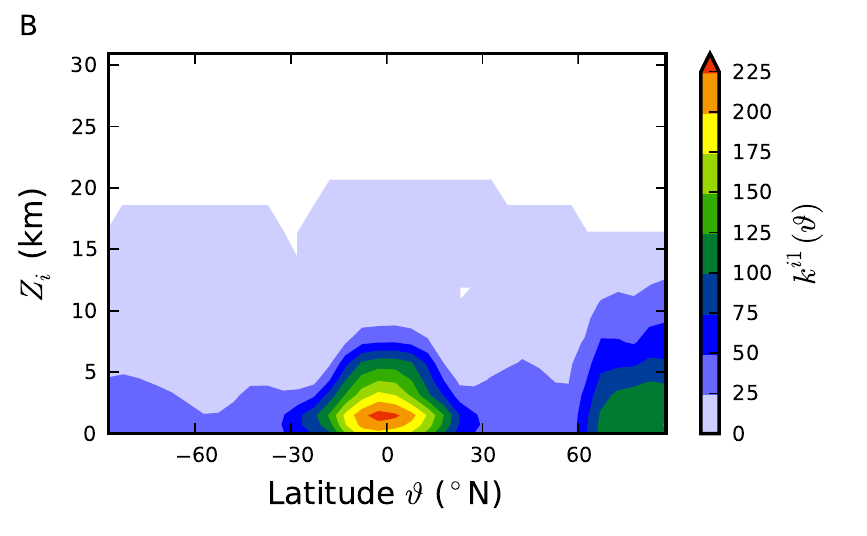}
}
\resizebox{1.0\textwidth}{!}{%
  \includegraphics{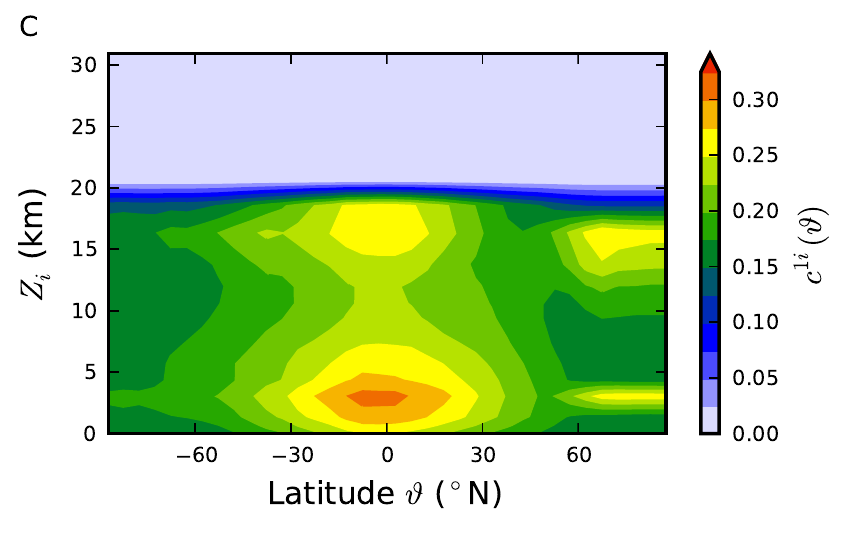}
  \includegraphics{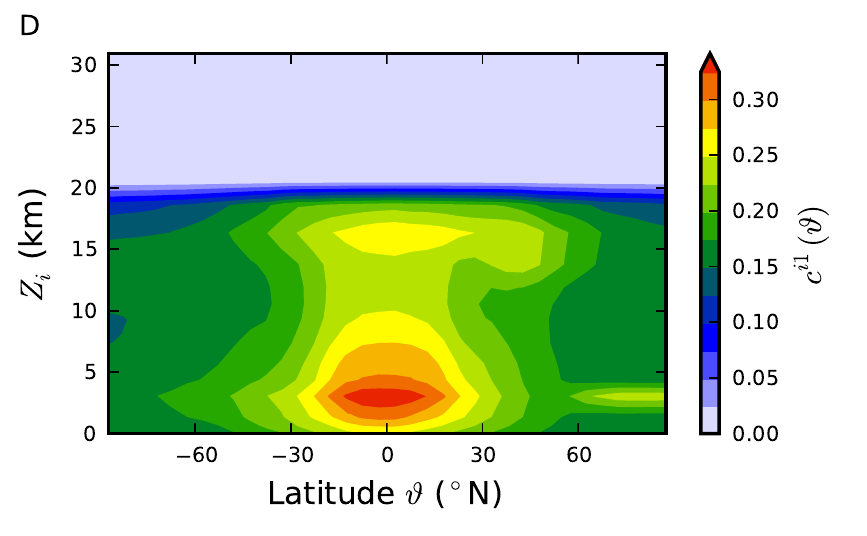}
}
\resizebox{1.0\textwidth}{!}{%
  \includegraphics{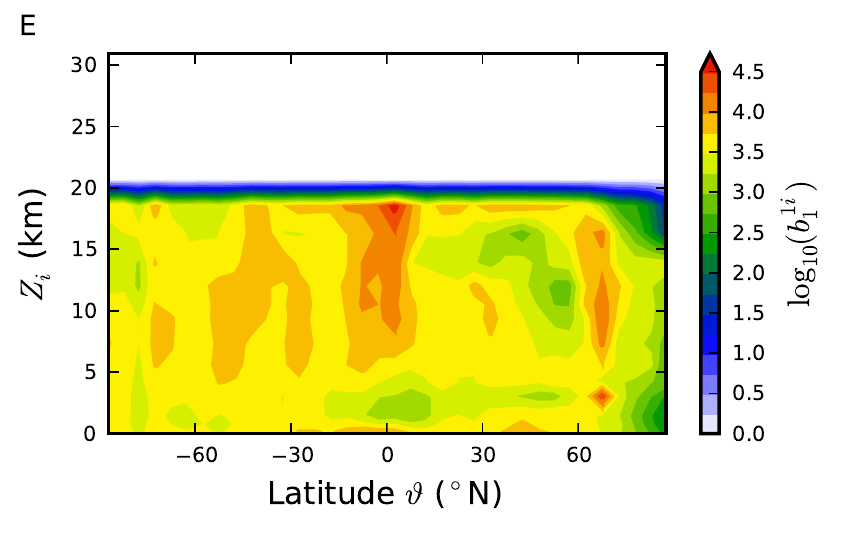}
  \includegraphics{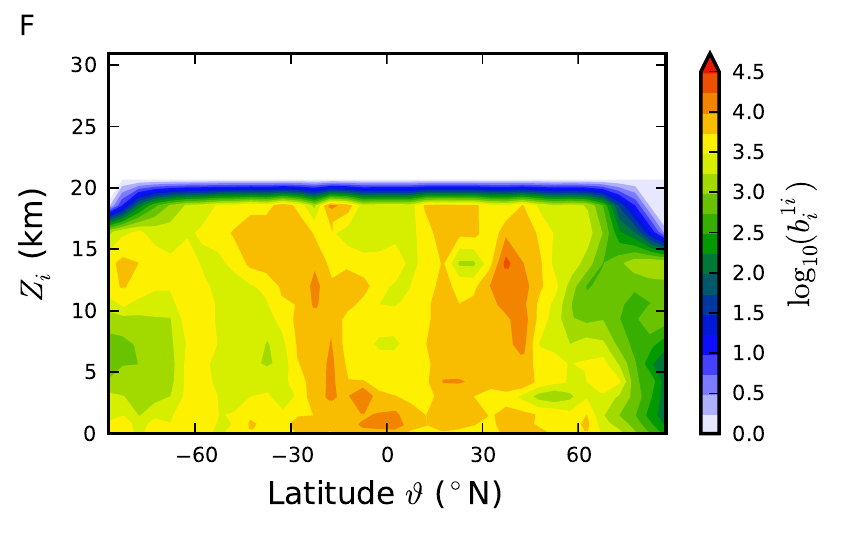}
}
\caption{As Fig. \ref{fig:cross_degree_closeness_vertical} with threshold $T=0.6$}
\label{fig:cross_degree_closeness_vertical_T_0_6}
\end{figure*}

\paragraph{Local measures}

For visualising the inherently three-dimen\-sional fields of local cross-network measures (one of the subnetwork indices $i,j$ indices being fixed as in our application) $m_v^{ij}=m_{v(\vartheta,\phi)}^{ij}$, where $\vartheta$ and $\phi$ denote latitude and longitude, we choose to focus on their variation with height and latitude. This is most appropriate as our aim is to study from a complex network perspective aspects of the atmosphere's general circulation \cite{Schneider2006,Hartmann2007}, the dominant forcing of which is the latitudinal variation of radiative solar forcing \cite{Salby1996}. Hence, in the following we will consider zonal averages 
\begin{equation}
m^{ij}(\vartheta)=\left<m_{v(\vartheta,\phi)}^{ij}\right>_{\phi}
\end{equation}
along circles of constant latitude. A detailed study and interpretation of the latitudinally and longitudinally resolved fields of cross-network measures will be the subject of future work. Like the scalar measures of cross-clustering discussed above, most local cross-network measures are non-symmetric with respect to interchanging subnetworks and, hence, intrinsically directional. Therefore we distinguish ``upward" cross-degree centrality $k^{1i}(\vartheta)$ from ``downward" cross-degree centrality $k^{i1}(\vartheta)$, and ``upward" cross-close\-ness centrality $c^{1i}(\vartheta)$ from ``downward" cross-close\-ness centrality $c^{i1}(\vartheta)$. For brevity here we present results for representative thresholds of $T=0.4, 0.5, 0.6$ only.

Cross-degree and cross-closeness centrality show a similar structure in both directions (Figs.~\ref{fig:cross_degree_closeness_vertical}, \ref{fig:cross_degree_closeness_vertical_T_0_5}, \ref{fig:cross_degree_closeness_vertical_T_0_6}A--D). Both measures are generally increased in the tropics and polar regions, whereas they take smaller values in the mid-latitudes. We observe a pronounced asymmetry between both hemispheres as the cross-degree and cross-close\-ness centralities in the northern polar regions are significantly larger than those above Antarctica and the surrounding Southern Ocean in the southern hemisphere. Furthermore, a structure of increased ``downward" cross-degree $k^{i1}(\vartheta)$ and -closeness centrality $c^{i1}(\vartheta)$ appears above the northern mid-latitudes but not above those of the southern hemisphere (Figs.~\ref{fig:cross_degree_closeness_vertical}, \ref{fig:cross_degree_closeness_vertical_T_0_5}, \ref{fig:cross_degree_closeness_vertical_T_0_6}B and D). 

Additionally considering the dependence of both local measures on geopotential height $Z_i$, the ``upward" pointing cross-degree $k^{1i}(\vartheta)$ and -closeness centralities $c^{1i}(\vartheta)$ possess two distinct tropical maxima centred around $1$\,km and $16$\,km, as well as two northern polar maxima at $3$\,km and $16$\,km (Figs.~\ref{fig:cross_degree_closeness_vertical}, \ref{fig:cross_degree_closeness_vertical_T_0_5}, \ref{fig:cross_degree_closeness_vertical_T_0_6}A and C). While $k^{1i}(\vartheta)$ decreases monotonously with height in the mid-latitudes and above the polar regions of the southern hemisphere, $c^{1i}(\vartheta)$ maintains its bimodal structure there. The ``downward" projecting cross-degree $k^{i1}(\vartheta)$ and -closeness centralities $c^{i1}(\vartheta)$ display two pronounced tropical maxima at $3$\,km and $16$\,km, and a maximum centred around $14$\,km in the northern mid-latitudes (Figs.~\ref{fig:cross_degree_closeness_vertical}, \ref{fig:cross_degree_closeness_vertical_T_0_5}, \ref{fig:cross_degree_closeness_vertical_T_0_6}B and D). Furthermore, $c^{i1}(\vartheta)$ reveals a maximum of topological closeness to the near ground isobaric surface at $1$\,km above the northern polar regions, while $k^{1i}(\vartheta)$ decreases monotonically with height there.

It is worth noting that the observed extrema in cross-degree are directly related to those in cross-edge density via Eq. (\ref{eq:cross_edge_density}) and extrema in cross-closeness have an equivalent association to those of cross-average path length (Eq. (\ref{eq:apl_closeness})). Moreover, we point out that plots of cross-degree centrality like those presented in this work may be used to draw conclusions on the main sources and destinations of cross-edges without relying on full three-dimensional visualisations of the coupled network structure. For example, consider the region of increased ``downward" cross-degree $k^{i1}(\vartheta)$ between $11$\,km and $18$\,km of geopotential height above the northern mid-latitudes (Fig.~\ref{fig:cross_degree_closeness_vertical}, \ref{fig:cross_degree_closeness_vertical_T_0_5}, \ref{fig:cross_degree_closeness_vertical_T_0_6}B). It implies that a considerably large number of cross-edges connect this region directly to the whole near ground isobaric surface. We learn where exactly on the near ground surface those cross-edges originate from by looking at the ``upward" cross-degree $k^{1i}(\vartheta)$. It measures how many and from where on the near ground surface cross-edges connect to some higher isobaric surface. Now the regions of increased $k^{1i}(\vartheta)$ between $11$\,km and $18$\,km above the tropics and northern polar regions imply that many cross-edges originating in the tropics and northern polar regions of the near ground surface project to isobaric surfaces between $11$\,km and $18$\,km (Fig.~\ref{fig:cross_degree_closeness_vertical}, \ref{fig:cross_degree_closeness_vertical_T_0_5}, \ref{fig:cross_degree_closeness_vertical_T_0_6}A). As these are the only major structures in this range of geopotential height, we may conclude that a significant number of cross-edges must link the tropical and northern polar near ground isobaric surface with the northern mid-latitudes' upper troposphere to lower stratosphere between $11$\,km and $18$\,km.

In contrast to the latter two local measures, cross-betweenness $b_w^{ij}$ with $w \in V_i \cup V_j$ is symmetric with respect to exchanging the involved subnetworks (see Eq. (\ref{eq:cross_betweenness})), but assigns a value to vertices of both subnetworks. Therefore we will in the following analyse zonally averaged fields of cross-betweenness
\begin{equation}
b_i^{ij}(\vartheta)=\left<b_{w(\vartheta,\phi)}^{ij}\right>_{\phi, w \in V_i}
\end{equation}
for vertices taken from a specific isobaric subnetwork $i$. It was shown earlier for climate networks constructed from surface air temperature data that betweenness centrality (Eq. (\ref{eq:betweenness})) yields additional information when compared to degree (Eq. (\ref{eq:degree})) and closeness centrality (Eq. (\ref{eq:closeness})) \cite{Donges2009}. Similarly, the near ground and higher isobaric surface components of cross-betweenness centrality $b_1^{1i}(\vartheta)$, $b_i^{1i}(\vartheta)$ reveal rich structures which are partially complementary to those seen in the zonally averaged fields of degree and closeness centrality (Figs.~\ref{fig:cross_degree_closeness_vertical}, \ref{fig:cross_degree_closeness_vertical_T_0_5}, \ref{fig:cross_degree_closeness_vertical_T_0_6}E and F). Both fields highlight how frequently certain regions on the two isobaric surfaces are traversed by shortest paths connecting the near ground to a higher isobaric surface. Now due to our network construction procedure (Sect.~\ref{sec:network_construction}), shortest paths correspond to sequences of strongly and significantly statistically interrelated pairs of time series with a minimum number of intermediate steps. Hence, it is conceivable to assume that to a first approximation a markedly increased cross-betweenness centrality indicates that a region is particularly important for mediating interactions between two isobaric surfaces, while the contrary is true for regions with significantly decreased cross-betweenness \cite{Donges2009}. 

The hemispherically asymmetric near ground component of cross-betweenness centrality $b_1^{1i}(\vartheta)$ reveals that the northern subpolar regions are exceptionally important for mediating interactions between the near ground and all considered heights ranging from the lower troposphere across the tropopause to the lower stratosphere (Figs.~\ref{fig:cross_degree_closeness_vertical}, \ref{fig:cross_degree_closeness_vertical_T_0_5}, \ref{fig:cross_degree_closeness_vertical_T_0_6}E). The near ground tropics and southern polar regions appear only to be relevant for coupling the near ground to isobaric surfaces in the upper troposphere and above. In contrast, the upper surface component of cross-betweenness $b_i^{1i}(\vartheta)$ possesses a more pronounced hemispherical symmetry (Figs.~\ref{fig:cross_degree_closeness_vertical}, \ref{fig:cross_degree_closeness_vertical_T_0_5}, \ref{fig:cross_degree_closeness_vertical_T_0_6}F). Most notable and stable for various thresholds $T$ are the two tongues of increased $b_i^{1i}(\vartheta)$ ranging from the near ground tropics to the mid-latitudes and subpolar regions in the lower stratosphere. These structures indicate that the latitude at which shortest paths connecting near ground and upper isobaric surfaces arrive in the upper surface tends to increase towards the poles for growing geopotential height $Z_i$ in both hemispheres. One should be aware that the structures detected in both components of the cross-betweenness field in the stratosphere as well as in all other local and global measures above $20$\,km of geopotential height should be treated with care. This is because only very few edges exist between the near ground and isobaric surfaces in these height levels (Fig.~\ref{fig:link_densities}B) and, hence, statistically expected false detections or omissions of cross-edges are likely to induce recognisable changes in the respective measures \cite{Bialonski2010}. To quantitatively assess the effects of small changes in the interaction topology between subnetworks, new types of significance tests based on network null models need to be developed in future work.

After describing the results of our analysis we would like to draw attention to the fact that the observed correlations in various measures revealed by qualitatively similar structures like the inversions at mostly three different height levels are not necessarily a direct consequence of the measure's definitions (see Sect.~\ref{sec:theory}), but point to a specific type of network structure. While correlations in different measures quantifying distinct aspects of network topology need not be present for general network structures, they are prevalent in many different types of real-world networks and network models \cite{Costa2005}. Particularly, these correlations are expected to arise in spatially embedded functional climate (and brain) networks like the coupled climate subnetworks considered in this work, since the increased probability of spatially close vertices to be connected imposes a substantial constraint on network topology \cite{Donges2009,Bialonski2010,Radebach2010}.

\subsection{Climatological interpretation}
\label{sec:climatological_interpretation}

We are now in a position to elaborate on the climatological implications of our coupled climate subnetwork analysis. First, recall that the coupled climate subnetworks were constructed from monthly averaged time series of geopotential height describing the dynamics of the atmosphere's quasi-geostrophic wind field on longer than monthly time scales. Variability on shorter time scales, e.g., synoptic scale weather systems, is included in the averages but does not appear explicitly in the time series. Therefore, we can indeed expect the coupled climate subnetworks to represent the climatological mean state of the atmosphere's three-dimensional correlation structure, excluding the direct effects of such weather phenomena with typical lifetimes of clearly less than one month.

The cross-network measures discussed in Sect.~\ref{sec:results} reveal for a wide range of thresholds $T$ aspects of the atmosphere's stratification and, more importantly, physical processes which couple the dynamics on different isobaric surfaces despite the strong buoyancy constraint imposed by vertical stratification. First it should be noted that in accordance with physical considerations and observations, all cross-network measures consistently indicate that most atmospheric dynamics takes place within the troposphere with comparatively weak coupling to the superjacent stratosphere \cite{Salby1996}. Steadily increasing (decreasing) from the near ground to reach a first maximum (minimum) at $1$\,km to $3$\,km of geopotential height, cross-edge density $\rho_{1i}$ and cross-average path length $\mathcal{L}_{1i}$ indicate that the near ground and higher isobaric surfaces become dynamically more densely interwoven when ascending from the planetary boundary layer below approx. $1$\,km into the lower free atmosphere (Figs.~\ref{fig:link_densities}B and \ref{fig:cross_apl}). Bearing in mind the spatial continuity of the geopotential height field, spatially close vertices are likely to be connected and to share common neighbours. This implies that the typical correlation radius of near ground vertices to higher isobaric surfaces and vice versa increases throughout the lower troposphere. This observation is consistent with the influence of the Earth's surface orography on atmospheric flow exponentially decreasing with height in the planetary boundary layer via the Ekman effect. Hand in hand with the prevalence of turbulence, the formation of long-range dynamical couplings is inhibited by this essentially frictional effect. In turn, within the less turbulent free atmosphere above approximately $1$\,km the wind field behaves quasi-geostrophically and allows for the long-range propagation of dynamical influences between the near ground and higher altitudes \cite{Salby1996}.

Within the cross-edge density (Fig.~\ref{fig:link_densities}B), we observe that the first maximum shifts from approximately $4$ km to $3$ km for higher thresholds. This detail can be credited to the decreasing height of the planetary boundary layer with latitude, the typical zone within which convection cells form. The prominent tropical convection processes are known to be more diffusive than those in the mid-latitudes and, hence, yield lower correlation values. Thus higher thresholds rather account for mid-latitude convection phenomena rejecting the relatively low correlated processes in the tropics. The prevalence of the latter, on the other hand, is reflected in lower thresholds.

Above approx. $3$\,km of geopotential height, $\rho_{1i}$ and $\mathcal{L}_{1i}$ decrease (increase) again until reaching a local minimum (maximum) between $11$\,km and $14$\,km (Figs.~\ref{fig:link_densities}B and \ref{fig:cross_apl}), indicating a dominating effect of a more stable vertical stratification acting to inhibit dynamical couplings between the now vertically more separated isobaric levels $1$ and $i$. This may be understood considering that turbulent vertical mixing is significantly less prevalent in the free atmosphere than it is in the planetary boundary layer and convection is suppressed by baroclinic adjustment forcing the atmosphere to interact horizontally. The second local maximum (minimum) of $\rho_{1i}$ and $\mathcal{L}_{1i}$ at $16$\,km highlights that the cumulative action of tropical penetrative convection processes (hot towers) reaching up to this height mediates markedly increased dynamical interrelationships between the near ground and isobaric surfaces close to the tropopause \cite{Riehl1958}. The tropical origin of this coupling is more readily seen in the fields of zonally averaged cross-degree and -closeness centralities (Fig.~\ref{fig:cross_degree_closeness_vertical}, \ref{fig:cross_degree_closeness_vertical_T_0_5}, \ref{fig:cross_degree_closeness_vertical_T_0_6}). When entering the stratosphere, quickly decreasing (increasing) $\rho_{1i}$ and $\mathcal{L}_{1i}$ indicate that influences reaching from the near ground to these heights are strongly inhibited by the dynamical barrier formed by the temperature inversion which marks the boundary between troposphere and stratosphere. This conclusion is further supported by the random-like interconnectivity structure revealed by the ``downward" pointing global cross-clustering coefficient $\mathcal{C}_{i1}$ within the stratosphere (Fig.~\ref{fig:cross_clustering}B). The behaviour of internal edge density $\rho_{i}$ is consistent with the foregoing argumentation in the troposphere, its striking increase above the tropopause reveals the spatially uniform dynamics of the statospheric wind field (Fig.~\ref{fig:link_densities}A) \cite{Salby1996}.

The zonally averaged fields of cross-degree, -close\-ness and -betweenness centrality uncover features of the atmosphere's general meridional circulation \cite{Schneider2006,Hartmann2007}. Most evident are the Hadley and polar cells which are indicated by markedly increased values of both measures in the tropics and polar regions (Figs.~\ref{fig:cross_degree_closeness_vertical}, \ref{fig:cross_degree_closeness_vertical_T_0_5}, \ref{fig:cross_degree_closeness_vertical_T_0_6}). Here the generally rising motion of air above the equator and subpolar latitudes couples surface wind dynamics to the upper troposphere, which becomes apparent in both components of zonally averaged cross-betweenness (Figs.~\ref{fig:cross_degree_closeness_vertical}, \ref{fig:cross_degree_closeness_vertical_T_0_5}, \ref{fig:cross_degree_closeness_vertical_T_0_6}E and F). The surface component of cross-betweenness may also be interpreted to show a signature of the northern hemisphere circumpolar vortex around a typical height of approximately $5$\,km which is known to induce vertical air motion and, hence, vertical dynamical coupling (Figs.~\ref{fig:cross_degree_closeness_vertical}, \ref{fig:cross_degree_closeness_vertical_T_0_5}, \ref{fig:cross_degree_closeness_vertical_T_0_6}E) \cite{Andrews1987}. Supporting this interpretation, a corresponding signature is not seen in the southern hemisphere which is consistent with the Antarctic ice shield inhibiting the formation of a polar vortex there. Cross-degree and -close\-ness centrality also show that the height of the tropopause decreases towards the poles by a slight poleward shift of their maximum values towards lower altitudes, as this dynamical barrier strongly inhibits the propagation of signals from the near ground to the stratosphere. The Ferrel circulation may be involved in forming the comparatively weak dynamical interrelationships between the near ground southern subtropics and mid-latitudes with isobaric surfaces lying in the upper troposphere and lower stratosphere (Figs.~\ref{fig:cross_degree_closeness_vertical}, \ref{fig:cross_degree_closeness_vertical_T_0_5}, \ref{fig:cross_degree_closeness_vertical_T_0_6}A and C). Similarly, the remarkable coupling of wind dynamics within the upper troposphere and lower stratosphere of the northern mid-latitudes with the near ground tropics and northern polar regions might be related to the northern Ferrel cell (Figs.~\ref{fig:cross_degree_closeness_vertical}, \ref{fig:cross_degree_closeness_vertical_T_0_5}, \ref{fig:cross_degree_closeness_vertical_T_0_6}B and D, see also the discussion in Sect.~\ref{sec:results}).

\subsection{Outlook}

The analysis performed in this section serves as an illustration of the potentials of our network-based approach for studying statistical interrelationships between different fields of climatological observables as well as arbitrary data with a similar structure. It was shown that the sequence of coupled climate subnetworks constructed from three-dimensional geopotential height data contains a lot of information on known features of the atmospheric general circulation and stratification \cite{Holton2004}. However, particularly the measure cross-betweenness centrality is readily interpretable in the context of coupled climate subnetworks (see above), but reveals interesting structures, which we cannot obviously relate to known features of the atmosphere's dynamical structure. This in turn indicates that coupled climate subnetworks in general and cross-betweenness centrality in particular have the potential to uncover previously unknown features of the atmosphere's general circulation and may prove useful in the future to shed light on so far open questions of atmospheric dynamics \cite{Schneider2006}, specifically those considering the response to climate change \cite{Lenton2008}. While it should be born in mind that the results of our study are subject to known deficiencies and limitations of the NCEP/NCAR Reanalysis 1 data, the variable geopotential height we analyse here is considered as one of the most reliable products of this reanalysis, since it is strongly determined by observations and, hence, less dependent on the particular model used for data assimilation \cite{Kistler2001}. However, we suggest that the results generated from several independent reanalysis data sets should be compared if definite climatological conclusions are to be drawn in future studies using coupled climate subnetwork analysis.

Additional information may be extracted from the available coupled climate subnetworks by investigating the spatially fully resolved fields of local cross-network measures or relying on further measures such as the local cross-clustering coefficient (which was not shown here for brevity). In a next step, network null models with a randomised interaction topology could be developed and used to assess the statistical significance of observed local and global cross-network measures. The simplest meaningful network null model of this type that was already discussed in the context of (local) cross-clustering and cross-transitivity (Sec. \ref{sec:global_measures}) can be constructed by fully randomising the interaction structure between two subnetworks while keeping the number of cross-edges fixed, e.g., by first deleting all cross-edges and subsequently redistributing them randomly between the two subnetworks. Furthermore, the network construction methodology may be fine-tuned, e.g., by using measures for detecting nonlinear or even directional interrelationships between time series or alternatively by including edges in the network based on the statistical significance of their associated correlation strengths.

Summarising the results of this first application of our framework for interacting network analysis, the particular advantage of this approach is that substantial conclusions can be drawn by analysing the dynamical correlation structure of the three-dimensional geopotential height field alone, without considering fields of temperature, moisture content or other relevant climatological variables. Subject to future work is the application of the interacting networks approach to fields of distinct climatological observables (e.g., surface air temperature and sea surface salinity) to further investigate the coupled dynamical behaviour of different components of the climate system.

%
%
\section{Conclusions}
\label{sec:conclusion}

In summary, we have developed a novel graph-theoretical framework for investigating in detail the interaction topology between pairs of subnetworks embedded within a network of networks. Applying this framework to analyse the correlation structure of a four-dimensional (spatio-temporal) data set of the climatological variable geopotential height yielded a consistent picture of the large scale circulation of the Earth's atmosphere. Particularly, the new measure cross-betweenness centrality was shown to have the potential to reveal previously unknown features of and to help address open questions on the atmosphere's general circulation \cite{Schneider2006}, particularly when considering its response to climate change \cite{Lenton2008}. Our results suggest that the coupled climate subnetwork approach presented in this work opens promising perspectives for the integrated analysis of several fields of climatological observables or, more generally, spatially embedded fields of arbitrary time series in the context of Earth system analysis. Particularly it will serve researchers as a tool complementary to established linear methods for the joint analysis of several climate data sets like canonical correlation analysis or singular value decomposition of the covariance matrix between two fields \cite{Bretherton1992}. Furthermore, we expect the proposed general graph-theoretical framework to meet the increasing need for investigating and understanding the interaction of complex systems from domains as diverse as social science, technology and engineering or the life sciences, as well as to stimulate further research in this direction, e.g., into supplementing the analysis with sophisticated null models designed for assessing selected aspects of interacting network structure.

\begin{acknowledgement}
This work has been financially supported by the Federal Ministry for Education and Research (BMBF) via the Potsdam Research Cluster for Georisk Analysis, Environmental Change and Sustainability (PROGRESS) and the Leibniz association (project ECONS). JFD thanks the German National Academic Foundation for financial support. The authors acknowledge the use of NCEP Reanalysis Derived data provided by the NOAA/OAR/\-ESRL PSD, Boulder, Colorado, USA, from their web site at \texttt{http://www.cdc.\-noaa.gov/} and are grateful to Jakob Runge for interpolating the data to an icosahedral grid. Thanks to John Scott McKechnie, Jobst Heitzig, Alexander Radebach, Reik Donner and Vladimir Petoukhov for valuable comments and discussions, and Karsten Kramer and Roger Grzondziel for help with the IBM iDataPlex Cluster at the Potsdam Institute for Climate Impact Research. Some graph theoretical calculations were performed using the software package \texttt{igraph} \cite{Csardi2006}.
\end{acknowledgement}

%
\bibliographystyle{epj}
\bibliography{library_epjb}

\begin{thebibliography}{59}

\bibitem{Albert2002}
R.~Albert, A.L. Barabasi, Rev. Mod. Phys. \textbf{74}, 47 (2002)

\bibitem{Newman2003}
M.E.J. Newman, SIAM Rev. \textbf{45}, 167 (2003)

\bibitem{Boccaletti2006}
S.~Boccaletti, V.~Latora, Y.~Moreno, M.~Chavez, D.U. Hwang, Phys. Rep.
  \textbf{424}, 175 (2006)

\bibitem{Costa2005}
F.A. Rodrigues, L.~da~F.~Costa, G.~Travieso, P.R. {Villas Boas}, Adv. Phys.
  \textbf{56}, 167 (2005)

\bibitem{Arenas2008}
A.~Arenas, A.~D\'{\i}az-Guilera, J.~Kurths, Y.~Moreno, C.~Zhou, Phys. Rep.
  \textbf{469}, 93 (2008)

\bibitem{Zhou2006}
C.S. Zhou, L.~Zemanov\'{a}, G.~Zamora-Lop\'{e}z, C.C. Hilgetag, J.~Kurths,
  Phys. Rev. Lett. \textbf{97}, 238103 (2006)

\bibitem{Zhou2007}
C.S. Zhou, L.~Zemanov\'{a}, G.~Zamora-L\'{o}pez, C.C. Hilgetag, J.~Kurths, New
  J. Phys. \textbf{9}, 178 (2007)

\bibitem{Barreto2008}
E.~Barreto, B.~Hunt, E.~Ott, P.~So, Phys. Rev. E \textbf{77}, 036107 (2008)

\bibitem{So2008}
P.~So, B.C. Cotton, E.~Barreto, Chaos \textbf{18}, 037114 (2008)

\bibitem{Vespignani2010}
A.~Vespignani, Nature \textbf{464}, 984 (2010)

\bibitem{Parshani2010}
R.~Parshani, S.~Buldyrev, S.~Havlin, Phys. Rev. Lett. \textbf{105}(4), 048701
  (2010)

\bibitem{Buldyrev2010}
S.V. Buldyrev, R.~Parshani, G.~Paul, H.E. Stanley, S.~Havlin, Nature
  \textbf{464}, 1025 (2010)

\bibitem{Fortunato2010}
S.~Fortunato, Phys. Rep. \textbf{486}, 75 (2010)

\bibitem{Kurant2006a}
M.~Kurant, P.~Thiran, Phys. Rev. Lett. \textbf{96}, 138701 (2006)

\bibitem{Kurant2006b}
M.~Kurant, P.~Thiran, Phys. Rev. E \textbf{74}, 036114 (2006)

\bibitem{Kurant2007}
M.~Kurant, P.~Thiran, P.~Hagmann, Phys. Rev. E \textbf{76}, 026103 (2007)

\bibitem{Donges2009}
J.F. Donges, Y.~Zou, N.~Marwan, J.~Kurths, Europhys. Lett. \textbf{87}, 48007
  (2009)

\bibitem{Donges2009a}
J.F. Donges, Y.~Zou, N.~Marwan, J.~Kurths, Eur. Phys. J. ST \textbf{174}, 157
  (2009)

\bibitem{Donner2008}
R.V. {Donner}, T.~{Sakamoto}, N.~{Tanizuka}, in \emph{{Nonlinear Time Series
  Analysis in the Geosciences: Applications in Climatology, Geodynamics and
  Solar-Terrestrial Physics}}, edited by R.V. Donner, S.M. Barbosa (Springer,
  2008), pp. 125--154

\bibitem{Gozolchiani2008}
A.~Gozolchiani, K.~Yamasaki, O.~Gazit, S.~Havlin, Europhys. Lett. \textbf{83},
  28005 (2008)

\bibitem{Tsonis2004}
A.A. Tsonis, P.J. Roebber, Physica A \textbf{333}, 497 (2004)

\bibitem{Tsonis2008a}
A.A. Tsonis, K.L. Swanson, Phys. Rev. Lett. \textbf{100}, 228502 (2008)

\bibitem{Tsonis2008b}
A.A. Tsonis, K.L. Swanson, G.~Wang, J. Climate \textbf{21}, 2990 (2008)

\bibitem{Yamasaki2008}
K.~Yamasaki, A.~Gozolchiani, S.~Havlin, Phys. Rev. Lett. \textbf{100}, 228501
  (2008)

\bibitem{Schellnhuber1999}
H.J. Schellnhuber, Nature \textbf{402}, C19 (1999)

\bibitem{Lenton2008}
T.M. Lenton, H.~Held, J.W. Hall, E.~Kriegler, W.~Lucht, S.~Rahmstorf, H.J.
  Schellnhuber, Proc. Natl. Acad. Sci. USA \textbf{105}(6), 1786 (2008)

\bibitem{Donner2009}
R.V. Donner, S.~Barbosa, J.~Kurths, N.~Marwan, Eur. Phys. J. ST
  \textbf{174}(1), 1 (2009)

\bibitem{Schneider2006}
T.~Schneider, Annu. Rev. Earth Planet. Sci. \textbf{34}(1), 655 (2006)

\bibitem{Hartmann2007}
D.~Hartmann, J. Meteor. Soc. Japan \textbf{85B}, 123 (2007)

\bibitem{Zamora2009}
G.~Zamora-L{\'o}pez, C.S. Zhou, J.~Kurths, Chaos \textbf{19}, 015117 (2009)

\bibitem{Zamora2010}
G.~Zamora-L{\'o}pez, C.S. Zhou, J.~Kurths, Front. Neuroinformatics \textbf{4},
  1 (2010)

\bibitem{Freeman1979}
L.C. Freeman, Soc. Networks \textbf{1}, 215 (1979)

\bibitem{Watts1998}
D.J. Watts, S.H. Strogatz, Nature \textbf{393}, 440 (1998)

\bibitem{Heitzig2010}
J.~Heitzig, J.F. Donges, Y.~Zou, N.~Marwan, J.~Kurths, in preparation  (2011)

\bibitem{Flom2004}
P.L. Flom, S.R. Friedman, S.~Strauss, A.~Neaigus, Connections \textbf{26}, 62
  (2004)

\bibitem{Brandes2008}
U.~Brandes, Soc. Networks \textbf{30}, 136 (2008)

\bibitem{Newman2004}
M.E.J. Newman, M.~Girvan, Phys. Rev. E \textbf{69}, 026113 (2004)

\bibitem{Kutzbach1967}
J.~Kutzbach, J. Appl. Meteorol. \textbf{6}(5), 791 (1967)

\bibitem{Wallace1981}
J.~Wallace, D.~Gutzler, Mon. Weather Rev. \textbf{109}(4), 784 (1981)

\bibitem{Vautard1989}
R.~Vautard, M.~Ghil, Physica D \textbf{35}(3), 395 (1989)

\bibitem{Mudelsee2010}
M.~Mudelsee, \emph{Climate Time Series Analysis: Classical Statistical and
  Bootstrap Methods}, Vol.~42 of \emph{Atmospheric and Oceanographic Sciences
  Library} (Springer, 2010)

\bibitem{Bretherton1992}
C.S. Bretherton, C.~Smith, J.M. Wallace, J. Climate \textbf{5}, 541 (1992)

\bibitem{Salby1996}
M.L. Salby, \emph{{Fundamentals of Atmospheric Physics}} (Academic Press San
  Diego, CA, 1996)

\bibitem{Pedlosky1979}
J.~Pedlosky, \emph{{Geophysical fluid dynamics}} (Springer, 1979)

\bibitem{Bialonski2010}
S.~Bialonski, M.T. Horstmann, K.~Lehnertz, Chaos \textbf{20}, 013134 (2010)

\bibitem{Donner2010a}
R.V. Donner, Y.~Zou, J.F. Donges, N.~Marwan, J.~Kurths, Phys. Rev. E
  \textbf{81}, 015101 (2010)

\bibitem{Donner2010b}
R.V. Donner, Y.~Zou, J.F. Donges, N.~Marwan, J.~Kurths, New J. Phys.
  \textbf{12}, 033025 (2010)

\bibitem{Marwan2009}
N.~Marwan, J.F. Donges, Y.~Zou, R.V. Donner, J.~Kurths, Phys. Lett. A
  \textbf{373}, 4246 (2009)

\bibitem{Kistler2001}
R.~Kistler, E.~Kalnay, W.~Collins, S.~Saha, G.~White, J.~Woollen, M.~Chelliah,
  W.~Ebisuzaki, M.~Kanamitsu, V.~Kousky et~al., B. Am. Meteorol. Soc.
  \textbf{82}, 247 (2001)

\bibitem{Heikes1995}
R.~Heikes, D.A. Randall, Mon. Weather Rev. \textbf{123}, 1862 (1995)

\bibitem{Jones1998}
P.W. Jones, T.~Diyision, \emph{{A user's guide for SCRIP: A spherical
  coordinate remapping and interpolation package}}, Los Alamos National
  Laboratory, Los Alamos, New Mexico, USA (1998)

\bibitem{Majewski2002}
D.~Majewski, D.~Liermann, P.~Prohl, B.~Ritter, M.~Buchhold, T.~Hanisch,
  G.~Paul, W.~Wergen, J.~Baumgardner, Mon. Weather Rev. \textbf{130}, 319
  (2002)

\bibitem{Kramer2009}
M.A. Kramer, U.T. Eden, S.S. Cash, E.D. Kolaczyk, Phys. Rev. E \textbf{79},
  061916 (2009)

\bibitem{Langford2001}
E.~Langford, N.~Schwertman, M.~Owens, Am. Stat. \textbf{55}(4), 322 (2001)

\bibitem{Radebach2010}
A.~Radebach, Master's thesis, Humboldt University Berlin (2010)

\bibitem{Riehl1958}
H.~Riehl, J.~Malkus, Geophysica \textbf{6}(3-4), 503 (1958)

\bibitem{Andrews1987}
D.~Andrews, J.~Holton, C.~Leovy, \emph{{Middle atmosphere dynamics}} (Academic
  Press San Diego, CA, 1987), ISBN 0120585766

\bibitem{Holton2004}
J.R. Holton, \emph{{An Introduction to Dynamic Meteorology}}, 4th~edn.
  (Elsevier, 2004)

\bibitem{Csardi2006}
G.~Cs\'ardi, T.~Nepusz, InterJournal \textbf{CX.18}, 1695 (2006)

\end{thebibliography}

\end{document}